\documentclass[a4paper, 10pt]{article}

\usepackage[usenames, dvipsnames, svgnames, table]{xcolor}
\usepackage{jcapmod}
\usepackage{verbatim}
\usepackage[toc, page]{appendix}
\usepackage[T1]{fontenc}
\usepackage[latin9]{inputenc}
\usepackage[normalem]{ulem}

\usepackage{xfrac}
\usepackage{float}
\usepackage{graphicx}
\usepackage{hyperref}
\usepackage{amsmath}
\usepackage{amssymb}
\usepackage{microtype}
\usepackage{amsbsy}
\usepackage{color}

\pdfpageheight\paperheight
\pdfpagewidth\paperwidth

\usepackage[textwidth=0.71\paperwidth, textheight=0.8\paperheight]{geometry}

\usepackage[capitalise]{cleveref}
\usepackage{breakurl}
\usepackage{csquotes}
\usepackage{lmodern}
\usepackage{bm}
\usepackage{dsfont}
\usepackage{subcaption}

\numberwithin{equation}{subsection}

\makeatletter

\g@addto@macro\bfseries{\boldmath}
\makeatother

\pdfpageheight\paperheight
\pdfpagewidth\paperwidth

\renewcommand*{\vec}[1]{\bm{#1}}

\newcommand{\reals}{\mathbb{R}}

\newcommand{\slabh}{\texorpdfstring{\ensuremath{\E{16}^{(\mathrm{h})}}}{E16h}} 
\newcommand{\slabi}{\texorpdfstring{\ensuremath{\E{16}^{(\mathrm{i})}}}{E16i}} 

\newcommand*{\E}[1]{\texorpdfstring{\ensuremath{E_{#1}}}{E#1}}

\newcommand{\Z}{\mathbb{Z}}
\renewcommand{\i}{\bm{i}}

\newcommand{\lmlpmp}{\ell m \ell' m'}

\newcolumntype{C}[1]{>{\centering\arraybackslash$}m{#1}<{$}}

\allowdisplaybreaks

\definecolor{lightgray}{gray}{0.9}
\definecolor{Amber}{rgb}{1.0, 0.75, 0.0}
\definecolor{blizzardblue}{rgb}{0.67, 0.9, 0.93}
\definecolor{burningsand}{RGB}{220, 148, 129}
\definecolor{burgundy}{rgb}{0.5, 0.0, 0.13}

\makeatletter
\newcommand\x@closefrown[3]{\mathrel{\ooalign{$#1\frown$\cr\kern-#3em\raisebox{-#2ex}{$#1-$}\hfil\cr\hfil\raisebox{-#2ex}{$#1-$}\kern-#3em\cr}}}
\newcommand\closefrown{\mathchoice
  {\x@closefrown{\displaystyle}{0.23}{0.02}}%
  {\x@closefrown{\textstyle}{0.23}{0.02}}%
  {\x@closefrown{\scriptstyle}{0.17}{0.01}}%
  {\x@closefrown{\scriptscriptstyle}{0.12}{0.01}}%
}
\newcommand\x@overcf[2]{\overset{\vphantom{x}\smash{\raisebox{-0.3ex}{$#1\closefrown$}}}{#2}}
\newcommand\overcf[1]{\mathpalette\x@overcf{#1}}
\makeatother

\makeatletter
\DeclareRobustCommand{\rcite}[1]{%
  \rcite@aux#1,\@nil{#1}%
}
\def\rcite@aux#1,#2\@nil#3{%
  \if\relax#2\relax
    Ref.~\cite{#3}%
  \else
    Refs.~\cite{#3}%
  \fi
}
\makeatother

\let\orgautoref\autoref
\renewcommand{\autoref}{%
    \def\equationautorefname{Eq.}%
    \def\figureautorefname{Fig.}%
    \def\sectionautorefname{Sec.}%
    \def\subsectionautorefname{Sec.}%
    \def\subsubsectionautorefname{Sec.}%
    \orgautoref
}

\subheader{IFT-UAM/CSIC-26-143}

\title{Cosmic topology. Part IId. Eigenmodes and correlation matrices of lens spaces}

\author[a, b]{Mikel Martin Barandiaran,}
\author[c]{Catherine Petretti,}
\author[d, e, f]{Stefano Anselmi,}
\author[g, h]{Deyan~P.~Mihaylov,}
\author[g]{Anna Negro,}
\author[g]{Glenn D. Starkman,}
\author[i, g]{Andrius Tamosiunas,}
\author[a, g]{Yashar Akrami,}
\author[a]{George Alestas,}
\author[a]{Javier Carr\'on Duque,}
\author[g]{Craig J. Copi,}
\author[j, g]{Fernando Cornet-Gomez,}
\author[k]{Linn Htat Lu,}
\author[k]{Andrew H. Jaffe,}
\author[l]{Arthur Kosowsky,}
\author[a]{Jos\'e Javier Ortega G\'omez,}
\author[m]{Ricardo~G.~Rodrigues,}
\author[g]{Amirhossein Samandar,}
\author[g]{Cynthia Trendafilova}

\collaboration{COMPACT Collaboration}
\collaborationImg{\includegraphics[width = 0.2\textwidth]{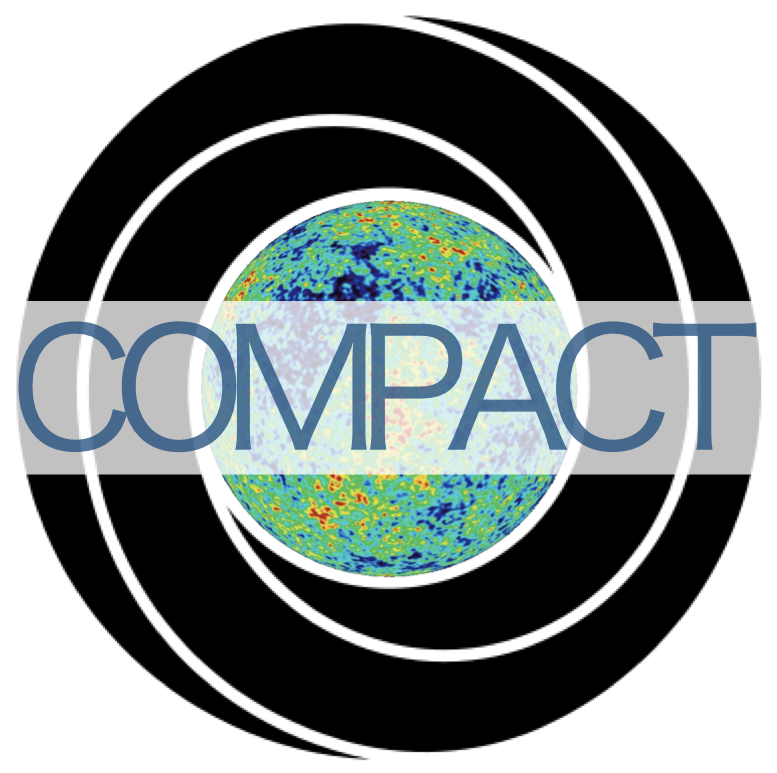}}

\affiliation[a]{Instituto de F\'isica Te\'orica (IFT) UAM-CSIC, C/ Nicol\'as Cabrera 13-15, Campus de Cantoblanco UAM, 28049 Madrid, Spain}
\affiliation[b]{Departamento de F\'isica Te\'orica, Universidad Aut\'onoma de Madrid, 28049 Madrid, Spain}
\affiliation[c]{Harvard-Smithsonian Center for Astrophysics (CfA), 60 Garden St., Cambridge, MA 02138, USA}
\affiliation[d]{INFN, Sezione di Padova, via Marzolo 8, I-35131 Padova, Italy}
\affiliation[e]{Dipartimento di Fisica e Astronomia ``G. Galilei'', Universit\`a degli Studi di Padova, via Marzolo 8, I-35131 Padova, Italy}
\affiliation[f]{Laboratoire Univers et Th\'eories, Observatoire de Paris, Universit\'e PSL, Universit\'e Paris Cit\'e, CNRS, F-92190 Meudon, France}
\affiliation[g]{CERCA/ISO, Department of Physics, Case Western Reserve University, 10900 Euclid Avenue, Cleveland, OH 44106, USA}
\affiliation[h]{Department of Astronomy, Faculty of Physics, Sofia University ``St Kliment Ohridski'', 5 James Bourchier Blvd, 1164 Sofia, Bulgaria}
\affiliation[i]{Institute of Theoretical Astrophysics, University of Oslo, P.O. Box 1029 Blindern, N-0315 Oslo, Norway}
\affiliation[j]{Departamento de F\'isica, Universidad de C\'ordoba, Campus Universitario de Rabanales, Ctra. N-IV Km. 396, E-14071 C\'ordoba, Spain}
\affiliation[k]{Astrophysics Group \& Imperial Centre for Inference and Cosmology, Department of Physics, Imperial College London, Blackett Laboratory, Prince Consort Road, London SW7 2AZ, United Kingdom}
\affiliation[l]{Department of Physics and Astronomy, University of Pittsburgh, Pittsburgh, PA 15260, USA}
\affiliation[m]{Departamento de F\'{i}sica, Universidade Estadual de Londrina, Rod. Celso Garcia Cid, Km 380, 86057-970, Londrina, Paran\'{a}, Brazil}

\emailAdd{mikel.martin@uam.es}
\emailAdd{catherine.petretti@cfa.harvard.edu}

\date{\today}

\abstract{
The global topology of the Universe is a longstanding open question. In this work, we examine the statistical signatures of a positively curved universe with a Friedmann--Lema{\^\i}tre--Robertson--Walker metric and the topology of a lens space $L(p,q)$. Since these manifolds are generally statistically anisotropic and inhomogeneous, their cosmic microwave background (CMB) covariance matrices contain non-zero off-diagonal entries that depend on observer location. We compute these full harmonic-space covariance matrices for scalar perturbations for arbitrary lens spaces and observer position. Using the Kullback--Leibler divergence, we assess the distinguishability of these spaces from a simply connected three-sphere with the same curvature. The results show that topological signatures can still be significant in a cosmic-variance-limited regime even when the length, $d_{\rm NC}$, of the shortest loop around the Universe through the observer exceeds the diameter, $d_{\rm LSS}$, of the last-scattering surface by up to $10\%$. This distinguishability depends on the curvature, the lens space and the observer position mainly through the ratio $d_{\rm NC}/d_{\rm LSS}$, which is set not only by the curvature radius but also by $p$, $q$ and the observer location. Therefore, for any value of the spatial curvature, however small, there are lens spaces and observer positions for which $0.985\,d_{\rm LSS}<d_{\rm NC}<1.1\,d_{\rm LSS}$. Such observers would find no matched circles in the CMB large enough to have been detected to date, yet the topology would remain potentially discoverable through its statistical signatures.
}

\keywords{cosmic topology, cosmic anomalies, statistical isotropy, cosmic microwave background, large-scale structure}

\begin{document}
\maketitle
\flushbottom

\section{Introduction}
\label{sec:intro}

One of the fundamental assumptions of the standard cosmological model is that on large scales, the Universe is well described  as a simply connected Friedmann--Lema{\^\i}tre--Robertson--Walker (FLRW) spacetime---an expanding homogeneous and isotropic space.
There are three possible FLRW geometries: positively curved (\(S^3\)), flat (\(E^3\)), and negatively curved (\(H^3\)).  
Not so long ago, it was widely regarded as plausible that the curvature contributions to the Friedmann equations governing the expansion of the FLRW Universe could be significant or even dominant, $\Omega_{\rm K}\simeq 0.7$.
More recently, by assuming the $\Lambda$CDM cosmological model, analyses of primary cosmic microwave background (CMB) anisotropy data indicate that the curvature parameter is much closer to zero, with $\vert\Omega_{\rm K}\vert \lesssim 0.05$~\cite{Planck:2018vyg}, with other analyses reporting (contested) evidence for a closed Universe \cite{DiValentino2020, Handley2021}.
Incorporating other ``late-time'' physics and data sets, even more stringent limits have been obtained, so that $\vert\Omega_{\rm K}\vert \lesssim 0.01$, or even $0.001$~\cite{Planck:2018vyg}. 

Although current observations do not provide a confirmed detection of non-zero curvature, there is no compelling theoretical reason to assume $\Omega_{\rm K}=0$ \cite{Anselmi:2022uvj}. Moreover, any robust detection of non-zero curvature would dramatically alter the landscape for inflationary theory, requiring an explanation of how the curvature scale was not inflated far beyond the current Hubble radius.

Curvature is also of significant interest for the study of the topology of the Universe (see \rcite{COMPACT:2026uzi} for a recent review).
The possible topologies of the curved homogeneous three-manifolds are entirely different than those of Euclidean three-space~\cite{COMPACT:2023rkp,COMPACT:2025adc} and, even when the curvature scale $R_c\simeq (3~\mathrm{Gpc}) ~h^{-1}\vert\Omega_{\rm K}\vert^{-1/2}$ is significantly larger than the Hubble scale $c/H_0$, the topological signatures may allow them to be distinguished from one another and from Euclidean manifolds.

In this paper, we turn our attention to the possibility that the Universe is a lens space, a particular class of quotients of the three-sphere. Every compact spherical manifold is obtained as a quotient of the three-sphere by a finite freely acting subgroup of its isometry group O(4). 
Considerable progress has been made in understanding the Laplacian eigenmodes of spherical manifolds and their implications for CMB anisotropies. 
Early work derived explicit eigenmodes for lens and prism spaces \cite{lehoucq2004eigenmodeslensprismspaces}, followed by general algorithms applicable to arbitrary spherical manifolds \cite{Lachieze-Rey:2005nyz}. 
These tools enabled detailed investigations of specific topologies, most notably the Poincar{\'e} dodecahedral space as a possible explanation for the observed suppression of large-angle CMB correlations \cite{Caillerie:2007gd,Aurich:2004fq}. 
Subsequent studies extended these analyses to lens spaces \cite{Uzan:2003ea,Aurich:2011wb,Aurich:2012sp} and more general double-action manifolds \cite{Aurich:2012yd,Aurich:2012jc}, while searches for matched circles have progressively constrained the parameter space of observable spherical topologies \cite{Aurich:2005hg,COMPACT:2024qni}.

Despite these advances, previous work has largely relied on summary statistics---most commonly the effective angular power spectrum $C_\ell^{\rm eff}$, i.e., the diagonal $\ell=\ell'$, $m=m'$ part of the harmonic covariance averaged over $m$---rather than the full harmonic covariance matrix. 
While analytic expressions for the covariance have been known for specific cases, their numerical evaluation has remained prohibitively expensive for generic lens spaces, preventing their use in practice. 
Moreover, the literature lacks a unified framework that simultaneously treats the eigenmode construction, observer dependence, basis transformations, and covariance matrix calculation in a consistent manner. 
The purpose of this work is to provide that unified and computationally useful framework for the study of scalar perturbations in arbitrary lens spaces. 
We derive the scalar Laplacian eigenmodes in a form well-suited to numerical implementation, obtain the corresponding selection rules, consistently incorporate arbitrary observer locations, and combine these ingredients to compute the full CMB harmonic correlation matrices, including their off-diagonal structure. 
These results provide the theoretical and computational foundation required for future Bayesian analyses of lens-space cosmologies using CMB temperature and polarization data, and we make our code publicly available on GitHub~\cite{S3Github}.

This paper is organized as follows. In \cref{sec:spherical_manifolds}, we review the geometry of the three-sphere and introduce lens spaces together with the coordinate systems and conventions used throughout the paper. 
In \cref{sec:eigenmodes}, we derive the scalar Laplacian eigenmodes, obtain the lens space selection rules, study the dependence on observer position, and present the resulting CMB correlation matrices. 
In \cref{sec:results}, we illustrate the properties of these correlation matrices and discuss their implications for cosmological analyses. 
Finally, \cref{sec:conclusions} summarizes our results and outlines future applications.

Throughout this work, we follow the conventions from \cite{Varshalovich:1988ifq} for Wigner D-matrices, spherical harmonics, and Clebsch-Gordan (CG) coefficients. 

\section{Spherical manifolds and lens spaces}
\label{sec:spherical_manifolds}

In this section, we first revisit fundamental aspects of the three-sphere, from geometrical, topological and cosmological perspectives. Then, we recall the general procedure to construct arbitrary manifolds with a positively curved FLRW metric. Finally, we focus on the properties of lens spaces---a specific class of spherical spaces---introducing the necessary definitions and conventions necessary to proceed with their study in \cref{sec:eigenmodes}.

\subsection{General considerations for the three-sphere}\label{ssec:general_considerations_s3}

Our intuition regarding spherical spaces is largely built upon our experience with the familiar two-dimensional sphere. The three-sphere, however, is fundamentally different from its lower-dimensional counterpart. The fact that the three-sphere is a compact Lie group, while the two-sphere is not, implies that the former admits structures with no two-dimensional analogue, making two-dimensional intuition often misleading.

For example, any orientation-preserving isometry of the two-sphere (i.e., any element of SO(3)) must have at least two fixed points (the ones defining the axis of rotation). However, a rotation of the three-sphere need not fix any points, and in fact, there are isometries of the three-sphere that move every point the same distance, much in the same way that translations move all points the same amount in Euclidean space. Such isometries are called \textit{Clifford translations} and their existence implies that it is possible to build nontrivial homogeneous  spherical manifolds.\footnote{The hyperbolic three-space $H^3$, by contrast, does not admit any Clifford translations.}

All spherical manifolds have a constant positive curvature $\kappa = 1/R_c^2$, where $R_c$ is the comoving curvature radius. 
In cosmological terms, such manifolds have a negative curvature-density parameter, 
\begin{equation}\label{eqn:omegak}
    \Omega_{\mathrm{K},0} = -\left(\frac{c/H_0}{R_c}\right)^2.
\end{equation}
In what follows, we will drop the 0 subscript from the curvature parameter, despite the fact that we exclusively refer to its present-day value throughout this work. A three-sphere with radius $R_c$ has volume $V=2\pi^2R_c^3$ and great circles of length $d=2\pi R_c$.

There are many ways to think of the three-sphere, each of which is better-suited for a different purpose. 
The most familiar way is probably as a submanifold of $\reals^4$,
\begin{equation}\label{eqn:s3-real}
    S^3=\{\bm{x}\in\reals^4\,:\bm{x}^T\bm{x}=x_1^2+x_2^2+x_3^2+x_4^2=R_c^2\}\,,
\end{equation}
where $R_c$ is the radius of the three-sphere. 
In a similar fashion, it can also be represented as a subset of $\mathbb{C}^2$, 
\begin{equation}\label{eqn:s3-complex}
    S^3=\{(z_1,z_2)\in\mathbb{C}^2\,:|z_1|^2+|z_2|^2=R_c^2\}\,.
\end{equation}
Neither of these representations makes the Lie group structure of the three-sphere particularly transparent. For most of the developments in this work, it is more convenient to identify the three-sphere with the sphere of quaternions of norm $R_c$,
\begin{equation}
\mathbb H_{R_c}=\{\mathbf g\in\mathbb H:\|\mathbf g\|=R_c\},
\end{equation}
which is simply a rescaled copy of the Lie group of unit quaternions,
\begin{equation}
\mathbb H_{R_c}=R_c\,\mathbb H_1,\qquad \mathbb H_1\simeq\mathrm{SU}(2).
\end{equation}
Since the group structure is defined only on unit quaternions, we implicitly work with the normalized coordinates $\mathbf g/R_c\in\mathbb H_1$ whenever quaternion multiplication is used. In this representation, points of the three-sphere correspond to unit quaternions and orientation-preserving isometries act naturally by left and right quaternion multiplication. A brief summary of the relevant quaternionic identities is given in Appendix~\ref{app:quaternions}.

\subsection{Coordinate systems}
\label{ssec:coords}

In this subsection, we define two coordinate systems for describing the three-sphere: the hyperspherical and toroidal coordinate systems. 
These will be useful for our ultimate goal of constructing CMB correlation matrices for three-dimensional spherical manifolds (\cref{sec:eigenmodes}).

The hyperspherical coordinates $(\psi, \theta, \phi)$ are the higher-dimensional analog of the familiar spherical coordinates. They are defined by 
\begin{equation}\label{eqn:hyper-defn}
    \begin{aligned}
        x&=R_c\,\sin\psi\sin\theta\cos\phi \,,  \\
        y&=R_c\,\sin\psi\sin\theta\sin\phi \,, \\
        z&=R_c\,\sin\psi\cos\theta \,, \\
        w&=R_c\,\cos\psi \,,
    \end{aligned}
\end{equation}
with $\psi,\theta\in[0,\pi]$ and $\phi\in [0,2\pi)$. For any fixed value of $\psi\in(0,\pi)$, $\theta$ and $\phi$ parameterize a two-sphere of radius $R_c\sin\psi$. 
The points with $\psi=0$ or $\psi=\pi$ are degenerate cases in which the two-sphere collapses to a point. 
The spatial metric in these coordinates is 
\begin{equation}\label{eqn:ds2-hyper}
    ds^2=R_c^2\left[d\psi^2+\sin^2\psi(d\theta^2+\sin^2\theta d\phi^2)\right] \,,
\end{equation}
which is one of the forms in which the spatial part of the FLRW metric is usually written in the cosmological literature. 
The volume element is
\begin{equation}\label{eqn:vol-hyper}
    dV=R_c^3\,\sin^2\psi\,\sin\theta \,d\psi\, d\theta \,d\phi\,.
\end{equation}

On the other hand, toroidal (also known as Hopf) coordinates $(\chi,\xi_+, \xi_-)$ are defined by 
\begin{equation}\label{eqn:tor-defn}
\begin{aligned}
    x&=R_c\,\cos\chi\cos\xi_+\,, \\
    y&=R_c\,\cos\chi\sin\xi_+\,, \\
    z&=R_c\,\sin\chi\cos\xi_-\,, \\
    w&=R_c\,\sin\chi\sin\xi_-\,,
\end{aligned}
\end{equation}
\begin{figure}[ht]
    \centering
    \includegraphics[width=0.4\linewidth]{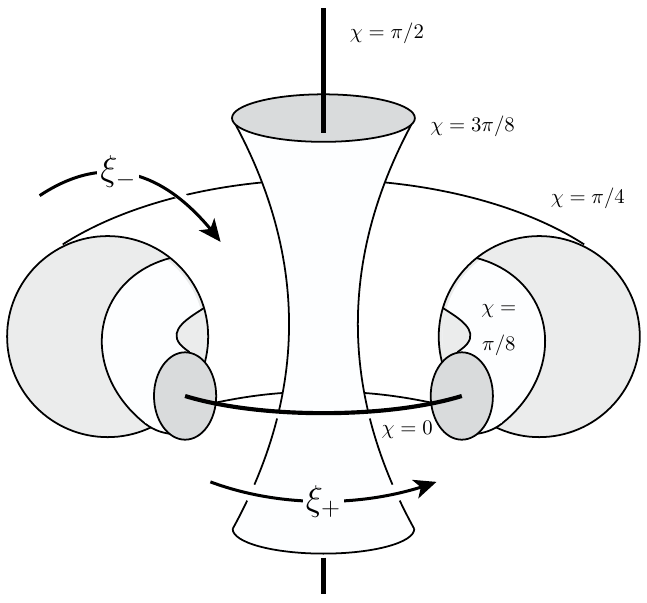}
    \caption{Toroidal coordinates foliate the three-sphere in two-dimensional tori, with two limiting great circles at $\chi=0$ and $\chi=\pi/2$. The figure shows the image of these tori under a stereographic projection of the three-sphere onto $\mathbb{R}^3$, which necessarily sends one point of the former (here on the $\chi=\pi/2$ circle) to infinity, so that circle appears as the vertical axis. Image adapted from \cite{lehoucq2004eigenmodeslensprismspaces}.}
    \label{fig:tor-coords}
\end{figure}
with $\chi\in[0,\pi/2]$ and $\xi_+,\xi_-\in[0,2\pi)$. 
For any fixed value of $\chi\in(0,\pi/2)$, $\xi_+$ and $\xi_-$ parameterize a flat two-torus of radii $R_c\cos(\chi)$ and $R_c\sin(\chi)$.
\Cref{fig:tor-coords} shows the image of these tori under a stereographic projection of the three-sphere onto $\mathbb{R}^3$.
The points with $\chi=0$ and $\chi=\pi/2$ are degenerate cases in which the two-torus collapses to a circle.

The spatial metric in these coordinates is
\begin{equation}\label{eqn:ds2-tor}
    ds^2=R_c^2\left[d\chi^2+\cos^2\chi d\xi_+^2+\sin^2\chi d\xi_-^2\right]\,,
\end{equation}
and the volume element is
\begin{equation}\label{eqn:vol-tor}
    dV=R_c^3\sin\chi\cos\chi\,d\chi\, d\xi_+ \, d\xi_- \,.
\end{equation}
The toroidal coordinates are related to the complex representation of the three-sphere (\cref{eqn:s3-complex}) via
\begin{equation}\label{eqn:complex_toroidal_relation}
    (z_1,z_2)=R_c(e^{i\xi_+}\cos\chi,e^{i\xi_-}\sin\chi)\in\mathbb{C}^2\,.
\end{equation}

Because the three-sphere is itself the Lie group SU(2), every point of the three-sphere can be regarded as a rotation, and hence labeled by a triple of Euler angles $(\alpha,\beta,\gamma)$---the standard coordinates on the rotation group. These are Euler angles of the point of the three-sphere itself, viewed as an element of SU(2), and they have nothing to do with the direction $(\theta,\phi)$ on an observer's sky. We introduce them here because the basis of Laplacian eigenmodes used throughout this work, built from the Wigner D-matrices of \cref{sec:eigenmodes-toroidal}, is conventionally written as a function of $(\alpha,\beta,\gamma)$. With the conventional ranges $\alpha,\gamma\in[0,2\pi)$ and $\beta\in[0,\pi]$, the triple covers SO(3)$\cong S^3/ \mathbb{Z}_2$ exactly once, and therefore only half of the three-sphere. Recovering all of the three-sphere requires doubling the range of one of the two azimuthal Euler angles, for instance $\gamma\in[0,4\pi)$, and it is the origin of the modular reductions below. Explicitly, the toroidal angles map to the usual SO(3) Euler angles in a 2-to-1 manner as
\begin{equation}\label{eqn:toroidal_euler_map}
    (\chi,\xi_+,\xi_-)\quad \longmapsto \quad (\alpha, \beta, \gamma) = \left(-(\xi_-+\xi_+)_{{\rm mod}~2\pi},\, 2\chi,\, (\xi_--\xi_+)_{{\rm mod}~2\pi} \right) \,.
\end{equation}
The modular reductions in \cref{eqn:toroidal_euler_map} are appropriate to the SO(3) reading of the angles, but they must \textit{not} be applied when the Euler angles are used as arguments of the eigenmodes of \cref{eqn:toroidal-eigenmodes}.

\subsection{Isometries and quotients of the three-sphere}
\label{ssec:isom_and_quotients}
All three-manifolds admitting a positively curved FLRW metric arise as quotients of the three-sphere by a discrete freely acting subgroup $\Gamma$ of SO$(4)$.\footnote{Although the isometry group of the three-sphere is technically O$(4)$, any isometry that has no fixed points is orientation-preserving, and therefore belongs to SO$(4)$. It follows that every spherical three-manifold is orientable.} 
The classification of all such subgroups was first completed in \cite{Threlfall1931}. 
The key to understanding the subgroups of interest and their classification is the following correspondence:
\begin{equation}
\mathrm{SO}(4)\cong (\mathrm{SU}(2)\times \mathrm{SU}(2))/\Z_2\cong (S^3\times S^3)/\Z_2\,.
\end{equation}
A very natural way to derive this correspondence is presented in Appendix~\ref{app:quaternions}, making use of the quaternionic point of view. The task of finding discrete subgroups of SO(4) is thus translated to the easier problem of considering the discrete subgroups of SU(2), together with the ways in which two such subgroups can be combined into a subgroup of the product. The compact nature of these groups, along with the discreteness condition, implies that all groups of interest will be finite. It then follows that the volume of any spherical three-manifold is 
\begin{equation}
\mathrm{Vol}(S^3/\Gamma)=\frac{\mathrm{Vol}(S^3)}{|\Gamma|}=\frac{2\pi^2R_c^3}{|\Gamma|}\,.
\end{equation}
An interesting remark, that we shall come back to later on, is that it is possible to construct spherical manifolds that are as flat as desired (i.e., with $R_c$ arbitrarily large) but remain small in terms of volume, simply by choosing a sufficiently large subgroup $\Gamma$.

Coming back to the types of spherical manifolds, the main result is that finite groups acting freely on the three-sphere can be organized into three broad classes, depending on the primary group $G$ appearing in the quotient group. In each case, the primary group can be supplemented by a cyclic factor $\mathbb{Z}_m$, so that the corresponding manifolds can be written as quotients of the form $S^3/(G\times \mathbb{Z}_m)$, with $G$ the primary group associated to each class. The precise way in which the two factors act distinguishes the single-, double-, and linked-action cases discussed below.

\begin{itemize}
\item \textbf{Lens spaces.} These are manifolds of the form $S^3/\Z_p$.\footnote{We do not bother writing $S^3/(\mathbb{Z}_n\times\mathbb{Z}_m)$ here: if $n$ and $m$ are coprime, $\mathbb{Z}_n\times\mathbb{Z}_m\cong\mathbb{Z}_{nm}$, and otherwise $\mathbb{Z}_n\times\mathbb{Z}_m$ cannot act freely on the three-sphere.} A more detailed account of lens spaces and their properties, which are the focus of this work, is given in \cref{ssec:lens_spaces}.
\item \textbf{Prism/dihedral spaces.} They arise as quotients of the type $S^3/(D_{4n}^*\times\mathbb{Z}_m)$, where $D_{4n}^*$ is the binary dihedral group of order $4n$. The word \textit{binary} here refers to the fact that it is the lift from SO(3) to SU(2) of the familiar dihedral group $D_{2n}$ comprised of the symmetries of a regular \textit{n}-gon. In the mathematical literature, it is slightly more common to denote the binary dihedral groups as Dic(n), and call them dicyclic groups.
\item \textbf{Binary polyhedral spaces.} These spaces are quotients of the three-sphere by double covers of the symmetry groups of the familiar Platonic solids. They can be divided in three subcategories:
\begin{itemize}
\item \textbf{Tetrahedral spaces.} These are quotients of the form $S^3/(T^*\times\mathbb{Z}_m)$, where $T^*$ is the binary tetrahedral group, of order 24, which is also isomorphic to SL(2,3).
\item \textbf{Octahedral spaces.} These are quotients of the form $S^3/(O^*\times\mathbb{Z}_m)$, where $O^*$ is the binary octahedral group, of order 48.
\item \textbf{Icosahedral spaces.} These are quotients of the form $S^3/(I^*\times\mathbb{Z}_m)$, where $I^*$ is the binary icosahedral group, of order 120, which is also isomorphic to SL(2,5). The Poincar\'e dodecahedral space is a special case of the icosahedral spaces where $m=1$, i.e., it is of the form $S^3/I^*$.
\end{itemize}
\end{itemize}

When $m=1$, one obtains the \textit{single-action} manifolds. When the cyclic factor acts independently of the primary group, subject to the appropriate coprimality condition, the resulting manifolds are referred to as \textit{double-action} spaces. Finally, \textit{linked-action} manifolds arise from suitable subgroups of the corresponding double-action groups in which the two factors are correlated. For this last case, the notation $S^3/(G\times \mathbb{Z}_m)$ is not fully precise, as the quotienting group is actually a subgroup of $G\times \mathbb{Z}_m$. The natural way to understand these distinct possibilities is again through the lens of quaternions and left/right-handed Clifford translations, but diving into the exact details is beyond the scope of this work. For a more comprehensive treatment of the subject, see \rcite{Gausmann:2001aa, Aurich:2012yd, Aurich:2012jc}.

\subsection{Lens spaces}
\label{ssec:lens_spaces}

Lens spaces, generically denoted as $L(p,q)$, constitute a well-studied family of spherical three-manifolds. 
They are constructed by taking the quotient of the three-sphere by the cyclic group $\mathbb{Z}_p$. 
The integer $p$ does not fully characterize the action of the isometries on the sphere; a second integer $q$ is needed. 
To produce a smooth manifold and avoid double-counting equivalent manifolds, $p$ and $q$ should be chosen coprime and in the range $0<q<p$.\footnote{The particular case $p=1$ corresponds to a quotient by the trivial group, which leaves the three-sphere untouched. Thus, it is customary to push the notation so that $L(1,1)\cong S^3$.}
Note that we retain both $q$ and $p-q$, rather than restricting to $0<q\leq p/2$ as is often done. The only value of $p$ for which $q=p-q$ is $p=2$, since otherwise $p=2q$ violates coprimality.

Choosing distinct allowed pairs $(p,q)$ and $(p',q')$ can nonetheless lead to topologically equivalent  (i.e., homeomorphic) manifolds. Specifically, $L(p,q)$ and $L(p,q')$ are homeomorphic if and only if $q'\equiv\pm q^{\pm1}\!\!\pmod p$,\footnote{Here $q^{-1}$ denotes the inverse of $q$ modulo $p$, i.e., the integer satisfying $q\,q^{-1}\equiv1\pmod p$, which exists because $p$ and $q$ are coprime.} and the homeomorphism can be chosen orientation preserving if and only if $q'\equiv q^{\pm1}\!\!\pmod p$. However, it was recently pointed out in \cite{COMPACT:2024qni} that some such pairs of homeomorphic lens spaces are physically distinguishable, due to the fact that they possess different handedness (chirality) in the pattern of clones. 

\begin{figure}[htbp]
    \centering
    \begin{minipage}{0.46\textwidth}
        \centering
        \includegraphics[width=\linewidth]{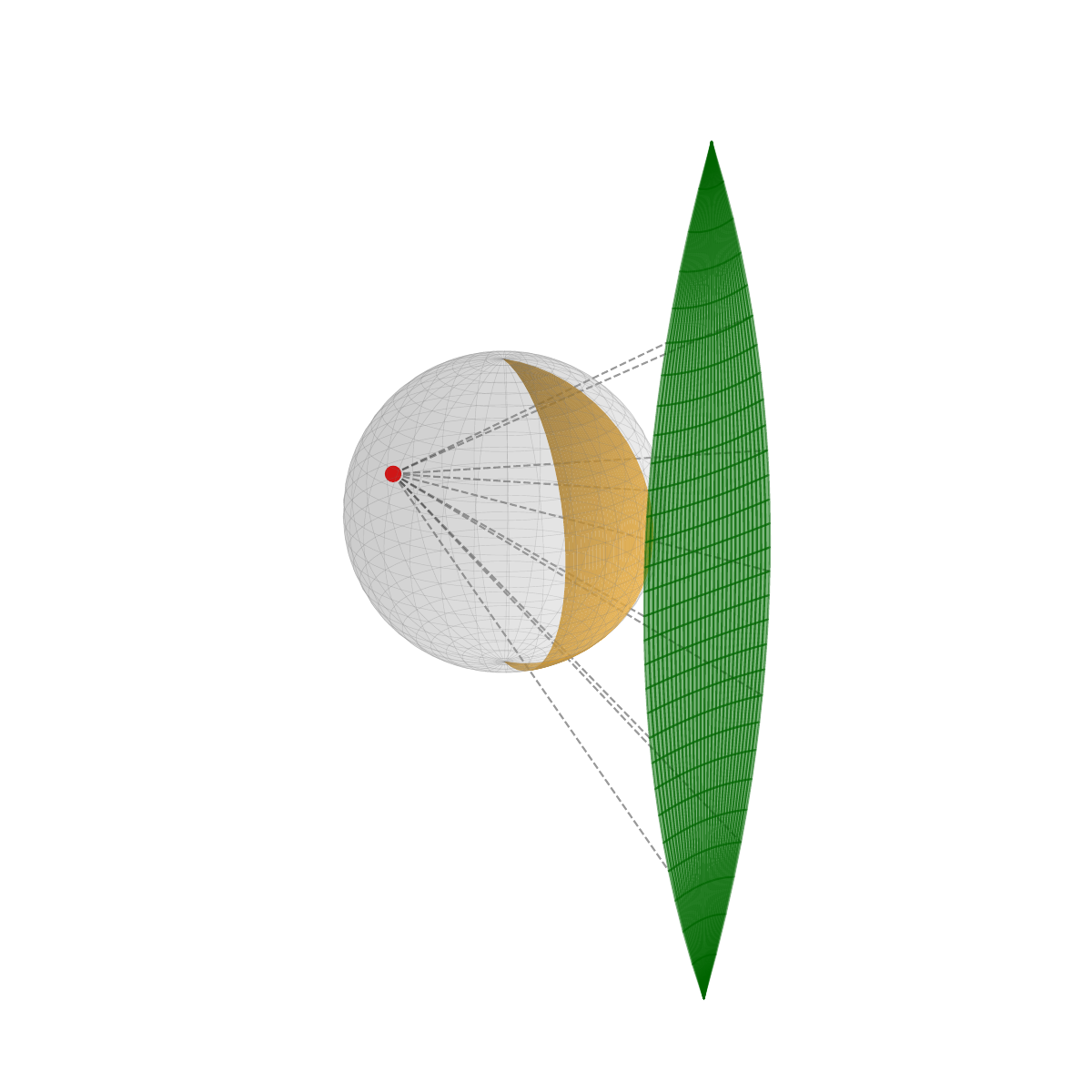}
    \end{minipage}\hfill
    \begin{minipage}{0.48\textwidth}
        \centering
        \includegraphics[width=\linewidth]{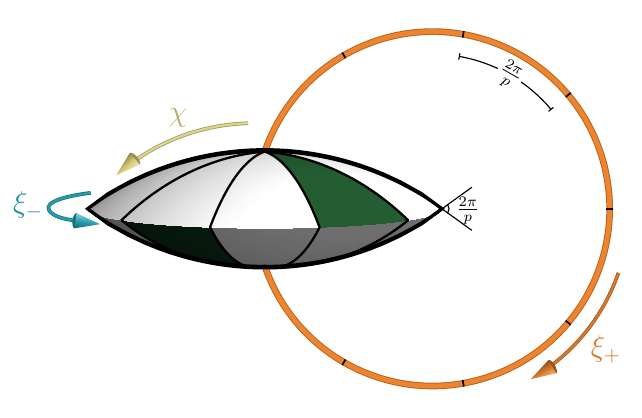}
    \end{minipage}
    \caption{ Visualization of the fundamental domain of a lens space obtained through stereographic projection. \textit{Left:} A wedge of the two-sphere is projected onto the plane, illustrating how a lens-shaped region naturally arises. \textit{Right:} The analogous construction for the three-sphere, where a wedge is stereographically projected into $\mathbb{R}^3$ to produce the lens-shaped fundamental domain of $L(p,q)$. Opposite curved faces are identified after a relative rotation determined by the parameter $q$. Adapted from Ref.~\cite{BALCH2021101712}.}
    \label{fig:lens_spaces}
\end{figure}

In the same way that a compact Euclidean manifold can be visualized as a Fundamental Domain (FD) with a set of boundary conditions that match faces of the FD, a lens space can be represented as a lens-shaped solid whose boundary faces are identified after an appropriate rotation.  
Since the three-sphere is naturally embedded in four-dimensional Euclidean space, such a picture is obtained through a stereographic projection from the three-sphere to $\mathbb{R}^3$. 
\Cref{fig:lens_spaces} illustrates this construction. 
The left panel shows the analogous, lower-dimensional situation in which a wedge of the two-sphere is stereographically projected onto the plane, producing the familiar lens-shaped region. 
The right panel depicts the corresponding FD for a lens space, obtained by stereographically projecting a wedge of the three-sphere into three-dimensional Euclidean space.
The opposite curved faces of this solid are then identified according to the action of the quotient group.
The volume of a lens space $L(p,q)$ with curvature radius $R_c$ is 
\begin{equation}
    \mathrm{Vol}(L(p,q))=2\pi^2R_c^3/p\,,
\end{equation}
which, as one would expect from \Cref{fig:lens_spaces},  is independent of $q$. This property also makes lens spaces interesting for quantum cosmology: the Euclidean action of the instanton that nucleates a closed universe scales with its spatial volume, and is therefore reduced by a factor $p$ for $L(p,q)$ at fixed curvature radius, at the price of the instanton being an orbifold instead of a manifold~\cite{Costa:1999ht,Fagundes:2006zy}. Depending on the boundary condition, this weakens the exponential suppression of the creation probability (tunneling) or favours the simply connected three-sphere (no-boundary)~\cite{PhysRevD.42.2458,Fagundes:2006zy,Hassfeld:2024hnx}. Revisiting these expectations for the large-$p$ lens spaces that remain observable at small $|\Omega_{\rm K}|$, in the spirit of the arguments made for compact flat and hyperbolic universes~\cite{Starobinsky:1993yx,Linde:2004nz}, would be an interesting direction for future work.
In the toroidal coordinate system introduced in \cref{ssec:coords}, the isometries that generate the lens space $L(p,q)$ act in a remarkably clean manner.\footnote{
    In the $\reals^4$ picture, the generator rotates the $x-y$ plane by $2\pi/p$ and the $z-w$ plane by $2\pi q/p$.
    Similarly, in the $\mathbb{C}^2$ picture, the rotation by $2\pi/p$ acts in the $z_1$ plane, and the rotation by $2\pi q/p$ acts in the $z_2$ plane.
} 
The generator of the quotient group $\mathbb{Z}_p$ leaves the radial-like angle $\chi$ strictly invariant, while inducing a discrete shift on the two azimuthal angles:
\begin{equation}\label{eqn:lens_isom}
    \chi \longrightarrow \chi\,, \quad
    \xi_+ \longrightarrow\xi_++\frac{2\pi}{p}\,,\quad
    \xi_- \longrightarrow\xi_-+\frac{2\pi q}{p}\,.
\end{equation}
Geometrically, points do not leave the fixed-$\chi$ 2-torus they belong to under the action of these isometries. 
In the limiting cases corresponding to $\chi\in\{0, \pi/2\}$ (that is, points lying in the great circles defining the azimuthal angles $\xi_+$ and $\xi_-$), the action of the generator is a simple rotation along the great circle.

While $L(p,1)$ and $L(p,p-1)$ manifolds are globally homogeneous, spaces with $1<q<p-1$ are inhomogeneous.
The distinction between homogeneous and inhomogeneous lens spaces can be seen straightforwardly by considering the distance to the \textit{nearest topological clone}. 
Given an observer at position $(\chi_0,{\xi_+}_0,{\xi_-}_0)$, each non-trivial element of the quotient group maps the observer to one of its images in the covering space. 
The geodesic distance to the nearest of these images defines the \textit{nearest-clone distance}, $d_{\rm NC}$, which sets the characteristic topological scale of the manifold.

For a lens space $L(p,q)$, the $j$-th non-trivial image ($j=1,\ldots,p-1$) is generated by applying Eq.~\eqref{eqn:lens_isom} $j$ times. The geodesic distance between a point and its $j$-th image is therefore
\begin{equation}\label{eqn:lens_clone_distance}
\cos\,\left(\frac{d_j}{R_c}\right)
= \cos^2\chi_0\,\cos\,\left(\frac{2\pi j}{p}\right) + \sin^2\chi_0\,\cos\,\left(\frac{2\pi jq}{p}\right)\,,
\end{equation}
which depends only on $\chi_0$.
The nearest-clone distance for an observer at $\chi_0$ is therefore 
\begin{equation}
d_{\rm NC}(\chi_0)=\min_{1\le j<p} d_j(\chi_0)\,.
\end{equation}
For the homogeneous spaces ($q=1$ and $q=p-1$, for which $\cos(2\pi jq/p)=\cos(2\pi j/p)$), \cref{eqn:lens_clone_distance} is independent of the observer position, yielding the simple expression
\begin{equation}
d_{\rm NC}=\frac{2\pi R_c}{p}\,.
\label{eqn:dnc_homogeneous}
\end{equation}
Thus, increasing $p$ reduces both the volume of the manifold and the distance to the nearest clone. 
By contrast, for inhomogeneous spaces ($q\neq1,p-1$), $d_{\rm NC}$ depends on the observer location through $\chi_0$. 
In this case one can define the maximum nearest-clone distance,
\begin{equation}
d_{\rm NC}^{\rm max}=\max_{\chi_0}\,d_{\rm NC}(\chi_0)\,,
\end{equation}
which corresponds to the observer position for which the topological signal is least apparent. 
This distance was used in Ref.~\cite{COMPACT:2024qni} to place constraints on the values of $(p,q)$ that are still observationally allowed, given the lack of matched circles in the CMB sky.

For cosmological applications it is convenient to compare $d_{\rm NC}$ with the diameter of the surface of last scattering, $d_{\rm LSS}$. The ratio $d_{\rm NC}/d_{\rm LSS}$ provides a natural measure of the observability of the topology: if $d_{\rm NC}\gg d_{\rm LSS}$, the observable Universe contains no clones and topological signatures are expected to be strongly suppressed, whereas values $d_{\rm NC}\sim d_{\rm LSS}$ allow for topological signatures to manifest. For the fiducial value of curvature adopted throughout this work, $\Omega_{\rm K}=-10^{-3}$, homogeneous lens spaces $L(p,1)$ with $p\simeq25$--$30$ satisfy
\begin{equation}
\frac{d_{\rm NC}}{d_{\rm LSS}}\simeq1.05\text{--}1.25\,,
\end{equation}
with $p$ = 30 giving the lower end of the range and $p$ = 25 the upper end, placing them close to the threshold where observable topological signatures are expected to become significant. This motivates the choice of representative models considered in the numerical analysis of ~\cref{sec:results}.

\section{Eigenmodes of the scalar Laplacian and correlation matrices of CMB scalar fluctuations}
\label{sec:eigenmodes}

In this section, we focus on the question of how to describe and study cosmological perturbations in a lens space. To that end, we first revisit the two natural bases of Laplacian eigenmodes that can be used to decompose fields and perturbations in a positively curved FLRW universe: the usual hyperspherical basis and the convenient toroidal basis. We then derive the selection rule for eigenmodes in the toroidal basis, as the isometries that lead to lens spaces are well adapted to toroidal coordinates, and specify how observer location affects the eigenmode expansion. Finally, by performing a change of basis back to the hyperspherical basis, we present the CMB harmonic correlation matrix $\langle a_{\ell m}a_{\ell^\prime m^\prime}^*\rangle\equiv C_{\ell m \ell^\prime m^\prime}$ for arbitrary observers in arbitrary lens spaces.

\subsection{The scalar Laplacian in the three-sphere}\label{ssec:laplacian}

The eigenvalues of the scalar Laplacian $\Delta$ in the three-sphere are of the form
\begin{equation}
  \text{Spec}(\Delta,S^3)=\{-q_n^2:q_n^2 = n(n+2)\kappa,\,n\in\mathbb{N}\} \,.
  \label{eqn:spectrum}
\end{equation}
The discreteness of the spectrum follows from the compactness of the three-sphere. 
The multiplicity of each eigenvalue $-q_n^2$, or equivalently, the dimension of the associated eigensubspace in the covering space, is $(n+1)^2$. As usual, we are interested in solving the eigenvalue problem
\begin{equation}\label{eqn:eigenvalueproblem}
    \Delta f=-q_n^2 f\,.
\end{equation}
Depending on which coordinates are used to write the explicit form of this equation, different families of eigenmodes are found. 
In toroidal coordinates 
\begin{equation}
    \Delta f = \kappa\left[\frac{1}{\sin\chi\cos\chi}\frac{\partial}{\partial\chi}\left(\sin\chi\cos\chi\frac{\partial f}{\partial\chi}\right) + \frac{1}{\cos^2\chi}\frac{\partial^2 f}{\partial\xi_+^2} + \frac{1}{\sin^2\chi}\frac{\partial^2 f}{\partial\xi_-^2}\right]\,;
\end{equation}
in hyperspherical coordinates,
\begin{equation}
    \Delta f = \kappa\left[\frac{1}{\sin^2\psi}\frac{\partial}{\partial\psi}\left(\sin^2\psi\frac{\partial f}{\partial\psi}\right) + \frac{1}{\sin^2\psi\,\sin\theta}\frac{\partial}{\partial\theta}\left(\sin\theta\frac{\partial f}{\partial\theta}\right) + \frac{1}{\sin^2\psi\,\sin^2\theta}\frac{\partial^2 f}{\partial\phi^2}\right]\,.
\end{equation}
Since we are ultimately interested in the CMB temperature and polarization fluctuations, which can be expanded as linear combinations of the eigenmodes of $\Delta$, it is useful to relabel the spectrum in terms of a wavenumber $k$ rather than the integer $n$. In flat space, the spatial frequency $k$ serves a dual purpose: its square ($-k^2$) is the Laplacian eigenvalue dictating the dynamical evolution of perturbations (the transfer functions), while $k$ itself acts as the geometric frequency in the radial wave equation. 

In the three-sphere, these two roles decouple. The dynamical equations governing the transfer functions rely directly on the eigenvalue itself, $q_n$. Instead, the geometric machinery relating the spatial eigenmodes to observable multipoles is built around the flat-space Helmholtz problem $\Delta f=-k^2 f$ in spherical coordinates. To reuse this geometric machinery the curved-space wavenumber $k$ must be the direct continuation of the flat-space spatial frequency, rather than merely the square root of the Laplacian eigenvalue $ -q_n^2$.

This can be seen by solving \cref{eqn:eigenvalueproblem} in hyperspherical coordinates $(\psi, \theta, \phi)$. Separating $f=\Phi(\psi)Y_{\ell m}(\theta,\phi)$ and substituting $u(\psi)\equiv\sin\psi\,\Phi(\psi)$ brings the radial equation to the form
\begin{equation}
    \frac{d^2u}{d\psi^2}+\left[(kR_c)^2-\frac{\ell(\ell+1)}{\sin^2\psi}\right]u=0\,,
    \label{eqn:radial}
\end{equation}
which is exactly the equation satisfied by $u=r\,j_\ell(kr)$ in flat space, with $R_c\psi\to r$ as $R_c\to\infty$. The radial functions $\Phi_{n\ell}(\psi)$ solving Eq.~\eqref{eqn:radial} are accordingly known as hyperspherical Bessel functions, reducing to the ordinary spherical Bessel functions $j_\ell(kr)$ in the flat limit~\cite{Harrison:1967zza,Abbott:1986ct,Kosowsky:1998nc,Lesgourgues:2013bra,Lewis:2026mif}. 

Compactness of the three-sphere imposes a second regularity condition for the solution at the antipodal point $\psi=\pi$, which has no flat-space analogue; together with regularity at $\psi=0$, it forces $kR_c$ to be a positive integer.\footnote{For $\ell=0$ this is immediate: Eq.~\eqref{eqn:radial} gives $u\propto\sin(kR_c\psi)$, and finiteness of $\Phi=u/\sin\psi$ at $\psi=\pi$ requires $\sin(kR_c\,\pi)=0$.} Writing $kR_c=n+1$, $n\in\mathbb{N}$, reproduces the spectrum of Eq.~\eqref{eqn:spectrum} and identifies the physically relevant wavenumbers as
\begin{equation}\label{eqn:kn_def}
    k_n=\frac{n+1}{R_c}\,,\quad n\in\mathbb{N}\,,
\end{equation}
in terms of which the eigenvalues take the compact form
\begin{equation}
    -q_n^2=-(k_n^2-\kappa)\,.
\end{equation}
The curvature offset $-\kappa$ is not a free convention: it is precisely the term generated by the curvature of the radial direction itself. We note that this set of discrete wavenumbers has a gap $\Delta k_n=1/R_c$, which tends to zero as  $R_c\rightarrow \infty$.\footnote{ In terms of observational sensitivity to the discreteness of $k_n$, in all realistic scenarios $R_c$ will be sufficiently large such that $k_n$ is effectively continuous.} 

In the cosmological literature there are several inequivalent notational conventions for these (generalized) wavenumbers that can cause confusion when crossing references. In particular, some sources swap the notations $k_n\leftrightarrow q_n$.

\subsection{The hyperspherical basis}\label{sec:eigenmodes-hyper}

As anticipated in the previous section, the eigenvalue problem \cref{eqn:eigenvalueproblem} in hyperspherical coordinates is most easily solved by performing separation of variables. Writing $\mathcal{Y}_{n\ell m}(\psi,\theta,\phi)=N\,R_{n\ell}(\psi)\,Y_{\ell m}(\theta,\phi)$ splits the eigenvalue equation into an angular piece, solved by the spherical harmonics
\begin{equation}
    Y_{\ell m}(\theta,\phi) = (-1)^m \sqrt{\frac{(2\ell+1)(\ell-m)!}{4\pi(\ell+m)!}}P_\ell^m(\cos\theta)e^{im\phi}\,,
\end{equation}
defined here with their usual normalization in terms of the associated Legendre polynomials $P_\ell^m(x)$, and a radial piece---\cref{eqn:radial}---solved by $R_{n\ell}(\psi)$. The indices are restricted to
\begin{equation}
    \{ 0\leq \ell \leq n\,, -\ell\leq m \leq \ell \,\vert\, n,\ell \in \mathbb{N},\, m\in\mathbb{Z}\}\,,
\end{equation}
with $N$ a normalization constant fixed below.
The radial eigenmodes can be written as
\begin{equation}\label{eqn:radial-eigenmodes}
    R_{n\ell}(\psi) = c_{n\ell}\,\sin^\ell \psi\, C_{n-\ell}^{(\ell+1)}(\cos\psi)\,, \quad  c_{n\ell}= 2^{\ell+1/2}\ell! \sqrt{\frac{(n-\ell)!(n+1)}{\pi (n+\ell+1)!}}\,,
\end{equation}
where $C_{n-\ell}^{(\ell+1)}(\cos\psi)$ is a Gegenbauer polynomial and the value of $c_{n \ell}$ comes from their orthonormality condition.

This is not the most common way of presenting the hyperspherical eigenmodes; the literature typically writes them instead as associated Legendre functions of  half-integer degree and order, multiplied by a square root of a trigonometric prefactor. That form makes the formal analogy with the negative-curvature case more transparent, but numerically the Gegenbauer form of \cref{eqn:radial-eigenmodes} is preferable, since orthogonal polynomials carry efficient recursion relations---a property exploited in Refs.~\cite{Tram:2013xpa,Lesgourgues:2013bra} to implement the transfer functions efficiently in the software package \texttt{CLASS}~\cite{Lesgourgues:2011re,Blas:2011rf}.

The overall normalization $N$ is not fixed by the eigenvalue equation itself, and different conventions coexist in the literature. We adopt the standard normalization of the hyperspherical Bessel functions, used in both \texttt{CAMB}~\cite{Lewis:1999bs,Lewis:2026mif} and \texttt{CLASS} (see also Refs.~\cite{Kosowsky:1998nc,Hu:1997mn}), choosing $N=\sqrt{\pi/2}/(n+1)$, so that
\begin{equation}
    N R_{n0}(\psi) = \frac{\sin[(n+1)\psi]}{(n+1)\sin \psi}\,,
\end{equation}
which fixes the hyperspherical eigenmodes to their final form,
\begin{equation}\label{eqn:eigenmodes-hyper}
    \mathcal{Y}_{n\ell m}(\psi,\theta,\phi) = \sqrt{\frac{\pi}{2}}\frac{1}{n+1} c_{n\ell}\sin^\ell \psi\, C_{n-\ell}^{(\ell+1)}(\cos\psi)\,Y_{\ell m}(\theta,\phi)\,.
\end{equation}
With this choice, the radial part of \cref{eqn:eigenmodes-hyper} is the hyperspherical Bessel function $\Phi^{\nu}_{\ell}(\psi)$ with $\nu=n+1$, in the Gegenbauer form of Ref.~\cite[Eq.~(2.4)]{Lesgourgues:2013bra}, and it tends to the flat radial modes $j_\ell(kr)$ for every $\ell$ in the $R_c\rightarrow\infty$ limit.

\subsection{The toroidal basis}\label{sec:eigenmodes-toroidal}

Similarly, we start by assuming separation of variables $\mathcal{T}(\chi,\xi_+,\xi_-)=A(\chi)B(\xi_+)C(\xi_-)$. Solving the eigenvalue equation in the toroidal basis yields
\begin{align} \label{eqn:toroidal-eigenmodes}
    {\mathcal T}_{nm_Lm_R}(\chi,\xi_+,\xi_-)&=\sqrt{\frac{n+1}{2\pi^2}}D^{n/2}_{m_Lm_R}(\xi_++\xi_-,2\chi,\xi_+-\xi_-)^* \\
    &= \sqrt{\frac{n+1}{2\pi^2}}D^{n/2}_{m_Lm_R}(-(\xi_++\xi_-),2\chi,\xi_--\xi_+)\, ,
\end{align}
where the asterisk indicates the complex conjugate. The allowed indices are
\begin{equation}\label{eqn:mLmR_defn}
    \left\{ n\in\mathbb{N},~ -\frac{n}2\leq m_L,m_R\leq \frac{n}2\right\}\,,
\end{equation}
where $m_L,m_R$ increase by unit steps, and can be either integers or half-integers depending on the value of $n$.
$D_{m_Lm_R}^{n/2}$ are the Wigner D-matrices, which are unitary irreducible representations of SU(2), defined as:
\begin{equation}\label{eqn:wignerD_def}
    D^{n/2}_{m_Lm_R}(\alpha,\beta,\gamma)=e^{-im_L\alpha}\,d^{n/2}_{m_Lm_R}\,(\beta)\,e^{-im_R\gamma}~,
\end{equation}
where $\alpha,\beta,\gamma$ are the Euler angles. The relation between the Euler angles and toroidal coordinates is given by \cref{eqn:toroidal_euler_map}. The Wigner small d-matrix is defined as 
\begin{equation}\label{eqn:dmatrix_def}
    d^{n/2}_{m_Lm_R}(\beta)=\sigma_{m_Lm_R}\sqrt{\frac{s!(s+\mu+\nu)!}{(s+\mu)!(s+\nu)!}}\left(\sin\frac{\beta}{2}\right)^\mu\left(\cos\frac{\beta}{2}\right)^\nu\,P^{(\mu,\nu)}_s(\cos\beta)\,,
\end{equation}
with
\begin{equation}
    \mu=|m_L-m_R|\,,\quad\nu=|m_L+m_R|\,,\quad s=\frac{n-\mu-\nu}{2}\,,\quad\sigma_{m_Lm_R}=(-1)^{(m_R-m_L-|m_R-m_L|)/2}\,,
\end{equation}
and $P^{(\mu,\nu)}_s(x)$ the Jacobi polynomials.

It is no coincidence that this basis of Laplacian eigenmodes is related to the Wigner D-matrices. This follows from the Peter-Weyl theorem, which states that for any compact Lie group $G$ the matrix coefficients of its irreducible representations form an orthonormal basis of $L^2(G)$. Therefore, since the three-sphere is a compact Lie group isomorphic to SU(2), the theorem applies directly: the matrix elements of the irreducible unitary representations of SU(2) (i.e., the Wigner D-matrices) furnish an orthonormal basis for $L^2(S^3)$.

\subsection{Change of basis}\label{ssec:change_basis}

For CMB computations we will be interested in writing the toroidal eigenmodes in the hyperspherical basis, as the $a_{\ell m}$ coefficients are most easily found by expanding formulae in terms of $Y_{\ell m}(\theta,\phi)$.

Note that for each eigenvalue, the associated eigenmodes form an orthonormal basis of the eigenspace. This means that the change of basis matrix $\langle\,n\,\ell\,m\,|\,n\,m_Lm_R\rangle$ must be unitary, which implies the following useful relations:
\begin{align}\label{eqn:unitarity_cond}
    \sum_{m_L,m_R}\langle\,n\,\ell\,m\,|\,n\,m_Lm_R\rangle\langle\,n\,\ell'\,m'\,|\,n\,m_Lm_R\rangle^*=\sum_{m_L,m_R}\langle\,n\,\ell\,m\,|\,n\,m_Lm_R\rangle\langle\,n\,m_Lm_R\,|\,n\,\ell'\,m'\rangle=\delta_{\ell\ell'}\delta_{mm'} ,\nonumber\\
    \sum_{\ell ,m}\langle\,n\,\ell\,m\,|\,n\,m_Lm_R\rangle^*\langle\,n\,\ell\,m\,|\,n\,m_L'm_R'\rangle=\sum_{\ell ,m}\langle\,n\,m_Lm_R|\,n\,\ell\,m\,\rangle\langle\,n\,\ell\,m\,|\,n\,m_L'm_R'\rangle=\delta_{m_Lm_L'}\delta_{m_Rm_R'}.
\end{align}
The inner product is diagonal in $m$ and $m_L+m_R$:
\begin{equation}\label{eqn:top_coeff_ortho}
    \langle\,n\,\ell\,m\,|\,n\,m_L\,m_R\rangle 
    \propto \int_0^{2\pi}d\phi\,Y_{\ell m}(\theta,\phi)^*e^{i(m_L+m_R)\xi_+} \propto \int_0^{2\pi}d\phi\,e^{i(m_L+m_R-m)\phi}=2\pi\,\delta_{m,m_L+m_R} ,
\end{equation}
where we have used that the hyperspherical azimuth coincides with the toroidal angle $\xi_+$, i.e., $\phi=\xi_+$. This means that an eigenmode in one of the bases is just a linear combination of a small subset of the eigenmodes in the other basis.  

To find the specific value of the change of basis coefficients, we can resort to representation theory. The $n$-th eigenspace of the Laplacian on the three-sphere forms an irreducible representation of SO(4). Since the double-cover of SO(4) is Spin(4) $\cong \text{SU}(2) \times \text{SU}(2)$, these eigenspaces lift to the $(n/2, n/2)$ irreducible representations of $\text{SU}(2) \times \text{SU}(2)$. In this framework, the toroidal eigenmodes (built from Wigner D-matrices) correspond to the uncoupled tensor product basis $|n/2, m_L\rangle \otimes |n/2, m_R\rangle$. However, for CMB applications, we require a basis adapted to the physical spatial rotation group SO(3). In the covering group, this corresponds to restricting the representation to the diagonal subgroup $\text{SU}(2)_{\text{diag}} \subset \text{SU}(2) \times \text{SU}(2)$. Under this restriction, the $(n/2, n/2)$ representation becomes reducible, decomposing into a direct sum of irreducible representations of total angular momentum $\ell$, where $0 \leq \ell \leq n$. The hyperspherical eigenmodes are precisely the basis states of these coupled representations. Therefore, the transformation between the toroidal (uncoupled) and hyperspherical (coupled) bases is fundamentally the quantum mechanical problem of adding two angular momenta, mediated directly by the Clebsch-Gordan (CG) coefficients (up to convention-dependent phases and normalizations) \cite{Stanton1990,Aurich:2012yd}:

\begin{equation}
    \langle\,n\,\ell\,m\,|\,n\,m_Lm_R\rangle \propto \left( \frac{n}{2}\,,m_L\,;\,\frac{n}{2},\,m_R\,|\,\ell ,m\right).
\end{equation}

To match the conventions set in \cite{Varshalovich:1988ifq}, the proportionality constant has to be the phase factor $(-1)^{m_L-|m_L|}\,i^{-n-\ell}$. Moreover, notice how CG coefficients automatically satisfy the requirements in \cref{eqn:unitarity_cond,eqn:top_coeff_ortho}.

\subsection{Lens space eigenmodes}

The Laplacian eigenmodes of a spherical manifold $S^3/\Gamma$ are precisely the eigenmodes of the three-sphere that are invariant under the isometries of $\Gamma$. As shown in \cref{eqn:lens_isom}, the isometries giving rise to lens spaces are well adapted to toroidal coordinates. Thus, it is easiest to find a basis of eigenmodes for the lens spaces by studying the selection rule that arises for the toroidal basis.\footnote{This is analogous to the case of the flat three-torus. The translation vectors defining the quotient act most naturally in Cartesian coordinates, not in spherical coordinates. Therefore, the natural eigenmode basis to retain is the plane wave basis $\{e^{i\bm{kx}}\}$, and the invariance condition (assuming orthogonal translation vectors) gives rise to the selection rule $\bm{k_n}=2\pi (n_x/L_x,n_y/L_y,n_z/L_z)$, $n_i\in\mathbb{Z}$, which comes from requiring the generated phase to equal unity.} Conveniently, when the isometry in \cref{eqn:lens_isom} acts on an eigenmode ${\mathcal T}_{nm_Lm_R}(\chi,\xi_+,\xi_-)$ only the exponentials in \cref{eqn:wignerD_def} are affected, and so the sole effect of the isometry is to produce an overall complex phase. By requiring that phase to equal unity, the selection rules for the eigensubspace indices $(m_L,m_R)$ are derived~\cite{lehoucq2004eigenmodeslensprismspaces}. In particular,
\begin{equation}\label{eqn:lens_invariance_condition}
  \exp\left(2\pi i\left[m_L\frac{1+q}{p}+m_R\frac{1-q}{p}\right]\right)=1\iff m_L(q+1)-m_R(q-1)\in p\Z  \,,
\end{equation}
which defines the subset of toroidal eigenmodes that furnish a basis $\mathcal{B}_{pq}$ of eigenmodes for $L(p,q)$:

\begin{equation}\label{eqn:lens_eigenmode}
\boxed{
    \mathcal{B}_{pq}=\{\mathcal{T}_{nm_Lm_R}(\chi,\xi_+,\xi_-)\,: n\in\mathbb{N}\,, (m_L,m_R)\in I^n_{pq}\}\,,  
    }
\end{equation}
where
\begin{equation}\label{eqn:selection_rule_lens}
\boxed{
I^n_{pq}=\{-n/2\leq m_L,m_R\leq n/2 : (q+1)\,m_L-(q-1)\,m_R\in p\,\Z\}~.  
}
\end{equation}
From \cref{eqn:selection_rule_lens}, it follows that for lens spaces with $(q=1)$, the invariant eigenmodes are those with $2m_L\in p\,\Z$. Notice that if $p$ is even, then $m_L$ has to be an integer, and so all eigenmodes with half-integer valued $m_L$'s (i.e., all eigenmodes with $n$ odd) are not allowed. In particular, for $L(2,1)$, also known as the elliptic space, the set of allowed indices is $m_L\in\Z\iff n\in 2\Z$. We thus see that all $m_L,m_R$ pairs associated with even $n$'s are allowed, and invoking \cref{eqn:unitarity_cond} it becomes clear why elliptic space is the only nontrivial FLRW topology that is globally isotropic and homogeneous.

For the computation of cosmological observables in lens spaces, the practical takeaway is that the eigenmode sums have to be restricted to eigenmodes with labels in $I^n_{pq}$. As a useful check, the number of retained pairs satisfies $|I^n_{pq}|\to(n+1)^2/p$ on average over $n$ as $n$ grows, reflecting the $p$-fold reduction in volume.

\subsection{Dependence on observer location}\label{ssec:location}

As anticipated in \cref{ssec:lens_spaces}, lens spaces with 
$1< q < p-1$
are statistically inhomogeneous---one observer sitting at coordinates $(\chi,\xi_+,\xi_-)$ and another observer sitting at $(\chi',\xi_+',\xi_-')$ need not measure the same statistical properties for the same observable. 
In order to capture the location dependence in the computation of observables, the origin of the coordinate system cannot be neglected in the eigenmode expansion.

For the toroidal eigenmodes, the effect of shifting the coordinate system is slightly more cumbersome than in 
Euclidean geometry. In the latter, transforming  $\bm{x}\rightarrow\bm{x}-\bm{x_0}$ makes the plane wave eigenmodes pick up a phase: $\exp(i\bm{kx})\rightarrow \exp(-i\bm{kx_0})\exp(i\bm{kx})$. 
In statistically homogeneous Euclidean manifolds, the extra phase $\exp(-i\bm{kx_0})$ cancels out naturally when computing observables, but it must be accounted for in the inhomogeneous manifolds. See \cite{COMPACT:2023rkp} for a complete treatment of the subject. 

In lens space manifolds, the effect of shifting the observer's location with respect to the natural axes that define the manifold is to make each toroidal eigenmode with eigenlabels in $I^n_{pq}$ a linear combination of \textit{all} toroidal eigenmodes of the same eigensubspace, even the ones that do not belong to $I^n_{pq}$. In other words, the restriction of $(m_L,m_R)$ to the index set $I^n_{pq}$ is only for the particular choices of coordinate origin that are on the natural axes of $L(p,q)$. The restriction to $I^n_{pq}$ should therefore be understood as a statement about the invariant basis in the natural coordinate system, rather than as a restriction on the labels appearing in an arbitrary observer-adapted expansion. The coefficients of the linear combination of eigenmodes that arise when moving the observer from the privileged origin follow from the fact that Wigner D-matrices form a representation of $\rm{SU}(2)\cong S^3$:
\begin{align}\label{eqn:wigner_expansion}
    D^{n/2}_{m_Lm_R}(\alpha',\beta',\gamma')=&\sum_{m_0}D^{n/2}_{m_Lm_0}(\alpha,\beta,\gamma)D^{n/2}_{m_Rm_0}(\alpha_0,\beta_0,\gamma_0)^*\\
    =&\sum_{m_0}D^{n/2}_{m_Lm_0}(\alpha,\beta,\gamma)D^{n/2}_{m_Rm_0}(-\alpha_0,\beta_0,-\gamma_0) ~,
\end{align}

which implies 

\begin{equation}\label{eqn:toroidal_translation}
    \mathcal{T}_{nm_Lm_R}(\chi',\xi_+',\xi_-')=
    \sum_{m_0}\,D^{n/2}_{m_Rm_0}({\xi_+}_0+{\xi_-}_0,\pi-2\chi_0,{\xi_+}_0-{\xi_-}_0)\mathcal{T}_{nm_Lm_0}(\chi,\xi_+,\xi_-) ~.
\end{equation}

In \cref{eqn:wigner_expansion,eqn:toroidal_translation}, the primed coordinates denote the shifted system, the unprimed coordinates the natural coordinates that define the lens space and the coordinates with a zero subscript denote the rotation angles that bring one to the other.\footnote{The relation between primed and unprimed coordinates is not as simple as in the flat case, where a subtraction of coordinates suffices. Instead, both coordinates are related to one another via the three-sphere's group operation. The cleanest way of seeing it is probably in terms of quaternions, which is expanded in Appendix \ref{app:quaternions}.} Equivalently, the zero subscript coordinates denote the position of the unshifted coordinate origin in the shifted system.

The second Euler angle in \cref{eqn:toroidal_translation} deserves a comment, since it is
$\pi-2\chi_0$ and not $2\chi_0$. The latter is the Euler angle of the observer's position
regarded as an element of SU(2) through \cref{eqn:toroidal_euler_map}, i.e., the angle
measured from the circle $\chi=0$. What is required here, however, is the rotation that
carries the origin of the hyperspherical chart onto the observer, and in light of \cref{eqn:hyper-defn,eqn:tor-defn}, the point $(0,0,0,R_c)$ corresponds to $\chi=\pi/2$. The two reference (great) circles are
separated by $\pi/2$ in $\chi$, hence by $\pi$ in the Euler angle $\beta$, which gives
$\beta=\pi-2\chi_0$.

\cref{eqn:toroidal_translation} defines the basis of eigenmodes that we shall use to compute CMB observables. We note that the mapping $(\chi,\xi_+,\xi_-)\rightarrow (\chi',\xi_+',\xi_-')$ that we have chosen is not technically a \textit{pure} translation, as it inherently contains an SO(3) rotation of the coordinate system beyond the one induced by parallel transport. 
In terms of quaternions, we choose to move the observer by applying a right Clifford translation $\mathbf{g}\rightarrow \mathbf{gu}$, instead of the geodesic translation given by $\mathbf{g}\rightarrow \sqrt{\mathbf{u}}\mathbf{g}\sqrt{\mathbf{u}}$, the two differing by the conjugation $\mathbf{g}\rightarrow\sqrt{\mathbf{u}}\,\mathbf{g}\,\sqrt{\mathbf{u}}^{-1}$, which is a rotation fixing the origin. 
The reasons to do so are twofold: on the one hand, the Kullback--Leibler (KL) divergence that we compute in \cref{ssec:kl} is invariant under rotations of the coordinate system; on the other hand, a mixed left and right Clifford translation mixes both $m_L$ and $m_R$ indices and therefore increases computational costs.
Moreover, rotations of harmonic two-point functions are best computed \textit{a posteriori}, with the simple rule that covariance matrices transform under an SO(3) rotation parameterized by Euler angles $(\alpha, \beta, \gamma)$ as
\begin{equation}
C_{\lmlpmp}\longrightarrow \sum_{\tilde m\tilde m'} D^\ell_{m\tilde m}(\alpha,\beta,\gamma)\, C_{\ell \tilde m \ell' \tilde m'} D^{\ell'}_{m'\tilde m'}(\alpha,\beta,\gamma)^*\,.
\end{equation}
\subsection{CMB correlation matrices}\label{ssec:corr_mats}

This subsection derives expressions for the matrices $C_{\ell m\ell' m'}^{XY}$, with $XY\in\{TT,TE,EE\}$, describing the correlations of CMB temperature $T$ and $E$-mode polarization harmonic coefficients induced by scalar fluctuations in a generic lens space $L(p,q)$:
\begin{equation}
C_{\ell m\ell' m'}^{XY;\,pq}\equiv\left\langle a^X_{\ell m}a^{Y*}_{\ell' m'} \right\rangle.
\end{equation}
We do not consider tensor perturbations in this work as this requires careful consideration of the spin-2 Laplacian (see, e.g., Ref.~\cite{COMPACT:2024cud,COMPACT:2025tqc}). We recall some basics of the three-sphere itself. Adopting the standard assumptions that
\begin{enumerate}
\item the amplitudes of eigenmodes describing a primordial cosmological field are statistically independent, zero-mean Gaussian random variables, and
\item due to statistical isotropy, the 2-point function of the eigenmode amplitudes is only a function of the eigenvalue label $n$ and not of the rest of the eigenlabels (the three-sphere analogue of $P(\bm{k})=P(k)$ in flat space),
\end{enumerate}
 it follows from \cref{eqn:unitarity_cond} that $\langle a^X_{\ell m}a^{Y*}_{\ell' m'}\rangle = C_\ell^{XY}\, \delta_{\ell\ell'}\delta_{mm'}$. The isotropy of the three-sphere therefore implies that the covariance matrix is diagonal in harmonic space and is completely specified by the angular power spectrum,
\begin{equation}\label{eqn:cl_s3}
    C_\ell^{XY;\,S^3}=\frac{2}{\pi R_c}\sum_{n=\ell}^{\infty}k_n^2 P^\mathcal{R}_{S^3}(k_n)
    \Delta_\ell^X(k_n) \Delta_\ell^Y(k_n)^*,   
\end{equation}
where the factor $1/R_c$ is the mode spacing $\Delta k_n$ of \cref{eqn:kn_def} replacing $dk$ in the flat space integral. Schematically,\footnote{For $E$-modes, $R_{n\ell}$ is replaced by the corresponding spin-2 radial function; see Ref.~\cite{Hu:1997mn}.
} transfer functions are of the form
\begin{equation}
    \Delta_\ell^X(k_n)= \sqrt{\frac{\pi}{2(n+1)^2}}\int d\tau  \, S^X(q_n,\tau)\, R_{n\ell}(\psi(\tau)),
\end{equation}
with $R_{n\ell}$ the hyperspherical radial eigenfunctions introduced in \cref{eqn:radial-eigenmodes} and $S^X$ the usual line-of-sight source functions, which we take from \texttt{CAMB}.
The source functions $S^X$ depend on the mode only through $q_n$, since $-q_n^2$ is the Laplacian eigenvalue entering the perturbation evolution equations. The radial eigenfunctions $R_{n\ell}$, which convert a spatial eigenmode into angular multipoles, are instead labeled by the geometric wavenumber $k_n$. In flat space the two coincide and the distinction is invisible, it is only in a spatially curved background that $\Delta_\ell^X$ carries both.
The dimensionful primordial spectrum $P^\mathcal{R}_{S^3}(k)$ is the curvature-corrected spectrum, and is described and discussed in \cref{ssec:primordial_ps}.

For a non-trivial spherical topology, the two assumptions stated above continue to hold for the primordial amplitudes of the allowed quotient eigenmodes. What changes is the completeness of the set of eigenmodes within each eigenspace. In the simply connected three-sphere, summing over the complete set of $(m_L,m_R)$ labels allows the unitarity relation in \cref{eqn:unitarity_cond} to be applied, yielding the Kronecker factors $\delta_{\ell\ell'}\delta_{mm'}$. In a quotient space, however, the sum is restricted to the subset $I^n_{pq}$ of modes that are invariant under the isometries giving rise to the quotient. The corresponding change-of-basis matrix is therefore no longer a complete unitary matrix on the covering-space eigenspace, and
\begin{equation}
\sum_{(m_L,m_R)\in I^n_{pq}}
\langle n\ell m|n m_Lm_R\rangle \langle n\ell' m'|n m_Lm_R\rangle^* \neq \delta_{\ell\ell'}\delta_{mm'}\,.
\end{equation}
It is this loss of completeness of the quotient eigenmode basis that allows correlations between different $(\ell,m)$ modes to arise. For a generic inhomogeneous lens space, the observer-dependent mixing further modifies these correlations and introduces an explicit dependence on the observer position. The consequence, thus, is the loss of statistical isotropy and (generally) homogeneity.

Combining the observer-dependent eigenmodes derived in \cref{ssec:location} with the transformation from the hyperspherical to the toroidal basis presented in \cref{ssec:change_basis}, the CMB covariance matrix for a lens space $L(p,q)$ reads
\begin{equation}
\label{eqn:clmlpmp_lens}
\boxed{
\begin{aligned}[b]
C_{\ell m\ell' m'}^{XY;\,pq}(\chi_0,{\xi_+}_0,{\xi_-}_0)
&=
\frac{2p}{\pi R_c} i^{\ell'-\ell} \sum_{n=\max\{\ell,\ell'\}}^\infty k_n^2 P^\mathcal{R}_{S^3}(k_n) \Delta_\ell^X(k_n) \Delta_{\ell'}^Y(k_n)^* \\
&\qquad\times \sum_{m_L,m_R\in I_{pq}^n} \left( \frac{n}{2},m_L; \frac{n}{2},m-m_L \middle| \ell m \right) \left( \frac{n}{2},m_L; \frac{n}{2},m'-m_L \middle| \ell' m' \right) \\
&\qquad\times D^{n/2}_{m_R,m-m_L}(\alpha_0,\beta_0,\gamma_0)\, D^{n/2\,*}_{m_R,m'-m_L} (\alpha_0,\beta_0,\gamma_0)\,,
\end{aligned}
}
\end{equation}
where
\begin{equation}
    \alpha_0 = {\xi_+}_0+{\xi_-}_0, \qquad \beta_0 = \pi-2\chi_0,\qquad \gamma_0 = {\xi_+}_0-{\xi_-}_0\,.
\end{equation}

The simply connected case $L(1,1)=S^3$ is recovered immediately, since the selection rules become trivial and \cref{eqn:clmlpmp_lens} reduces to the diagonal expression \eqref{eqn:cl_s3}.

A significant simplification occurs for observers located along the great circles defined by either $\chi_0=0$ or $\chi_0=\pi/2$. In these cases the reduced Wigner matrix satisfies
\begin{equation}
d^{n/2}_{m_R,m-m_L}(0)=\delta_{m_R,m-m_L}\,,\qquad d^{n/2}_{m_R,m-m_L}(\pi)=(-1)^{n/2-m+m_L}\,\delta_{m_R,m_L-m},
\end{equation}
which greatly simplifies the sum over $I_{pq}^n$. While the full Wigner D-matrices do pick up an overall complex phase that depends on $({\xi_+}_0,{\xi_-}_0)$, this phase perfectly cancels out in the product $D \times D^*$, so that the covariance matrix becomes
\begin{align}
\left. C_{\ell m\ell' m'}^{XY;\,pq} \right|_{\chi_0=\pi/2}&=
\frac{2p}{\pi R_c} i^{\ell'-\ell} \sum_{n=\max\{\ell,\ell'\}}^\infty k_n^2 P^\mathcal{R}_{S^3}(k_n) \Delta_\ell^X(k_n) \Delta_{\ell'}^Y(k_n)^* \nonumber\\
& \qquad\times \sum_{m_L,m_R\in I_{pq}^n} \left( \frac{n}{2},m_L; \frac{n}{2},m_R \middle| \ell m \right) \left( \frac{n}{2},m_L; \frac{n}{2},m_R \middle| \ell' m' \right).
\end{align}
The surviving terms are those with $m=m'=m_L+m_R$. The observer on the other great circle,
$\chi_0=0$, is obtained by replacing $m_R\rightarrow-m_R$ in both CG
coefficients, so that the selection becomes $m=m'=m_L-m_R$.

\subsection{The primordial power spectrum}
\label{ssec:primordial_ps}

Even in simply connected $\Lambda$CDM universes with $\Omega_{\rm K}\neq 0$, the form of the primordial power spectrum on scales approaching the curvature radius is not uniquely fixed, and different proposals exist~\cite{Handley:2019anl,Kiefer:2021iko,Vardanyan:2023ldj}.
A non-trivial topology introduces a similar freedom at the topology scale, as on scales comparable to $d_{\rm NC}$, the primordial statistics depend on the initial state of the perturbations and on whatever physics set the topology, neither of which is fixed by the standard inflationary paradigm. The spectrum on these scales is therefore a modeling choice rather than a prediction. Phenomenological departures from a power law on such scales have been studied in Ref.~\cite{COMPACT:2026pqs}, and other early-Universe consequences of a non-trivial topology in Refs.~\cite{COMPACT:2026vdj,Negro:2026ooh}.
Here we make the minimal choice of assigning to the eigenmodes allowed by the topology the same spectrum as in the covering three-sphere, so that a lens space and the three-sphere differ only through the selection of eigenmodes in \cref{eqn:selection_rule_lens} (see also the discussion at the end of \cref{ssec:kl}). Below we review the conventional form of the three-sphere spectrum, as adopted in cosmological software such as \texttt{CAMB}, which we use for our computations.

For adiabatic scalar perturbations, which are the ones we are interested in, the relevant initial conditions are set by the power spectrum of the primordial comoving curvature perturbation $\mathcal{R}$. In flat space this is defined through the two-point function
\begin{equation}
    \langle \mathcal{R}(\bm{k})\mathcal{R}^*(\bm{k}')\rangle = (2\pi)^3\,P^\mathcal{R}(k)\,\delta^{(D)}(\bm{k}-\bm{k}')\,,
\end{equation}
together with the associated dimensionless spectrum $\mathcal{P}^\mathcal{R}(k)\equiv k^3 P^\mathcal{R}(k)/2\pi^2$, normalized so that $\mathcal{P}^\mathcal{R}(k)\,d\ln k$ gives the contribution of a logarithmic interval around $k$ to the total variance of $\mathcal{R}$. For a nearly scale-invariant, slow-roll inflationary spectrum, this takes the familiar power-law form 
\begin{equation}
    \mathcal{P}^\mathcal{R}(k)=A_s\left(\frac{k}{k_{\rm piv}}\right)^{n_s-1}\,.
    \label{eqn:pps-flat}
\end{equation}

In a closed universe, \cref{eqn:pps-flat} cannot be applied directly to the discrete wavenumbers $k_n$ of \cref{eqn:kn_def}. The reason is that the wavenumber entering the mode equation for $\mathcal{R}$ during inflation---and in which the spectrum therefore takes the simple form \eqref{eqn:pps-flat}---is not $k_n$ itself, but
\begin{equation}
   q_n = \sqrt{k_n^2-\kappa}\,.
    \label{eq:qn_def}
\end{equation}
This is simply the curved-space version of the flat substitution $-\nabla^2\to k^2$. In the conventional treatment, the evolution equation for $\mathcal{R}$ during inflation is obtained from its flat-space counterpart by $k^2\to-\Delta$, so it is $q_n$, not $k_n$, that plays the role of the flat wavenumber there. Accordingly, both \texttt{CAMB}, which we use, and \texttt{CLASS} take the input primordial spectrum to be a function of $q_n$~\cite[Sec.~3.4]{Lesgourgues:2013bra}. Since $k_n$ and $q_n$ genuinely differ already at the level of the mode equation, and not merely by relabelling, the spectrum \eqref{eqn:pps-flat} must first be understood as a function of $q$, and only converted to a function of $k$ at the end---consistently with, and equivalent to, the wavenumber shift used by \cite[Eq.~(10)]{Hu:1997mn} for scalar perturbations, once their notation (in which the roles of $k$ and $q$ are swapped relative to ours) is translated into the conventions of this work.\footnote{We stress that the $q$ used in the context of the primordial power spectrum is not the same as the $q$ used to label the lens spaces $L(p,q)$. However, we retain using $q$ and $q_n$ as defined in \cref{eq:qn_def} to highlight and call attention to the common notation for the primordial power spectrum used by other works.
}

Since $\mathcal{P^R}(q)$ is, by construction, the contribution per logarithmic interval of $q$, converting it into a spectrum labeled by $k$ requires preserving this logarithmic weighting,
\begin{equation}
    \mathcal{P}^\mathcal{R}_{S^3}(k)\,\frac{dk}{k} \equiv \mathcal{P}^\mathcal{R}(q)\,\frac{dq}{q}\,.
\end{equation}
Differentiating \cref{eq:qn_def} gives $q\,dq=k\,dk$, and therefore
\begin{equation}
    \mathcal{P}^\mathcal{R}_{S^3}(k)=\frac{k^2}{q^2}\,\mathcal{P}^\mathcal{R}(q) = \frac{k^2}{k^2-\kappa}\,\mathcal{P}^\mathcal{R}\,\left(\sqrt{k^2-\kappa}\right) = \frac{k^2}{k^2-\kappa}\,A_s\left(\frac{\sqrt{k^2-\kappa}}{k_{\rm piv}}\right)^{n_s-1}\,,
    \label{eqn:pps-s3}
\end{equation}
in agreement with \cite[Eq.~(3.29)]{Lesgourgues:2013bra}. As a check, \cref{eqn:pps-s3} reduces to \cref{eqn:pps-flat} in the flat limit $\kappa\to0$, as it should.

The dimensionful spectrum entering \cref{eqn:cl_s3,eqn:clmlpmp_lens} follows from the same relation used in flat space,
\begin{equation}
    P_{S^3}^\mathcal{R}(k)\equiv \frac{2\pi^2}{k^3}\,\mathcal{P}_{S^3}^\mathcal{R}(k)\,,
\end{equation}
since this relation is simply a statement about the normalization of the dimensionless spectrum at a given value of $k$, and holds irrespective of whether $k$ is subsequently integrated over a continuum or summed over the discrete set $\{k_n\}$; no separate curvature correction is required at this step. The curvature-corrected spectrum $\mathcal{P}^\mathcal{R}_{S^3}(k_n)$ of \cref{eqn:pps-s3}---not the naive flat power law \eqref{eqn:pps-flat} evaluated at $k_n$---is therefore the quantity that must be used throughout \cref{eqn:cl_s3,eqn:clmlpmp_lens}.

\section{Numerical results}
\label{sec:results}

In this section, we present the numerical implementation of the formalism developed in the previous sections and discuss its application to the computation of CMB covariance matrices in lens spaces. We then use these covariance matrices to quantify the statistical distinguishability between lens-space cosmologies and the simply connected three-sphere through the Kullback--Leibler (KL) divergence~\cite{kullback1951,kullback1959information}.
Throughout this section the analysis is idealized: we assume full-sky, noiseless, cosmic-variance-limited observations of the temperature and $E$-mode fields, with no mask and no lensing. The resulting divergences should therefore be read as upper bounds on what any real experiment can extract.

\subsection{Evaluation of covariance matrices}
\label{ssec:eval_cov_mat}

Deriving closed expressions for the CMB covariance matrices constitutes only the first step towards constraining lens-space cosmologies with observations. In practice, the covariance matrices must also be evaluated accurately and efficiently over a large region of parameter space, making the numerical implementation equally important.

The main computational challenge originates from the loss of statistical isotropy. 
Unlike the standard $\Lambda$CDM covariance matrix, which is diagonal in harmonic space, the covariance matrices of generic lens spaces are fully populated---so that the number of independent entries grows as $\mathcal{O}(\ell_{\rm max}^4)$ rather than $\mathcal{O}(\ell_{\rm max})$---and therefore computations to evaluate them or that use them scale much less favourably with $\ell_{\rm max}$. 
In addition, evaluating Eq.~\eqref{eqn:clmlpmp_lens} requires computing CG coefficients and Wigner D-matrices for angular momenta as large as $n\gtrsim5000$ for the multipole range considered here.

A second practical difficulty is that most publicly available implementations of angular-momentum algebra are either designed for integer angular momenta only (while lens spaces naturally require half-integer values of $m_L$ and $m_R$) or become extremely slow and inaccurate for higher values of angular momenta. Therefore, standard libraries cannot be directly employed for the evaluation of Eq.~\eqref{eqn:clmlpmp_lens}.

For the CG coefficients, we use the exact prime-factorization algorithm \texttt{wigxjpf} developed in Ref.~\cite{Johansson:2015cca}, which remains numerically stable for the large angular momenta required in our calculations. For the Wigner D-matrices, we implement an independent routine, included in our publicly available code, based on the three-term recurrence relations of the Jacobi polynomials appearing in Eq.~\eqref{eqn:dmatrix_def}. This approach is both numerically stable and significantly faster than direct evaluation from the defining expressions. We validate the implementation by verifying row and column orthonormality of $d^{n/2}(\beta)$ to machine precision at the largest $n$ used, and by direct comparison of matrix entries with the values found in Ref.~\cite{Varshalovich:1988ifq}. Moreover, we have further validated our pipeline by verifying that the $L(1,1)$ case produces the usual $\Lambda$CDM diagonal power spectrum, matching the \texttt{CAMB} output  $C_\ell$'s within the tolerance of \cref{eqn:nmax_def} and with vanishing off-diagonal entries to machine precision.

In practice, the sum over $n$ in \cref{eqn:clmlpmp_lens} has to be truncated at some $n_{\rm max}$. To make this explicit, we write the covariance as a sum of contributions,
\begin{equation}\label{eqn:clmlpmp_summand}
    C_{\ell m\ell' m'}^{XY;\,pq}=\sum_{n=\max\{\ell,\ell'\}}^{\infty}\mathcal{C}_{\ell m\ell' m'}^{XY;\,pq}(n)\,,
\end{equation}
where $\mathcal{C}_{\ell m\ell' m'}^{XY;\,pq}(n)$ is the $n$-th term of \cref{eqn:clmlpmp_lens}, including the prefactor $2p\,i^{\ell'-\ell}/(\pi R_c)$ and the inner sum over $(m_L,m_R)\in I^n_{pq}$. For $p=q=1$, this inner sum reduces to $\delta_{\ell\ell'}\delta_{mm'}$ separately for each $n$, and the summand becomes $\delta_{\ell\ell'}\delta_{mm'}\,\mathcal{C}_{\ell}^{XY;\,S^3}(n)$, where
\begin{equation}\label{eqn:cl_s3_summand}
    \mathcal{C}_{\ell}^{XY;\,S^3}(n)\equiv\frac{2}{\pi R_c}\,k_n^2\,P^\mathcal{R}_{S^3}(k_n)\,\Delta_\ell^X(k_n)\,\Delta_\ell^Y(k_n)^*
\end{equation}
is the summand of \cref{eqn:cl_s3}. We take $n_{\rm max}$ to be the smallest integer for which the truncated three-sphere temperature spectrum agrees with the \texttt{CAMB} output to within $0.5\%$ at every multipole considered,
\begin{equation}\label{eqn:nmax_def}
    \max_{2\leq\ell\leq\ell_{\rm max}}\frac{\left|\sum_{n=\ell}^{n_{\rm max}}\mathcal{C}_{\ell}^{TT;\,S^3}(n)-C_\ell^{\texttt{CAMB}}\right|}{C_\ell^{\texttt{CAMB}}}\leq0.005\,,
\end{equation}
and truncate \cref{eqn:clmlpmp_summand} at this $n_{\rm max}$ for all $XY$ and all $(\ell m,\ell' m')$. Whenever a lens-space covariance matrix is compared with that of the three-sphere, both are truncated at the same $n_{\rm max}$, so that no difference between them arises from the number of retained terms. In particular, the truncated $L(1,1)$ matrix coincides exactly with the truncated three-sphere one. For $\ell_{\rm max}=30$ and $\Omega_{\rm K}=-10^{-3}$, this gives $n_{\rm max}\simeq4000$. The required $n_{\rm max}$ increases with $\ell_{\rm max}$ and, at fixed $\ell_{\rm max}$, scales approximately as $R_c\propto|\Omega_{\rm K}|^{-1/2}$ since convergence is reached at a physical wavenumber $k_{n_{\rm max}}=(n_{\rm max}+1)/R_c$ that depends only weakly on curvature.

The structure of \cref{eqn:clmlpmp_lens} determines where the computational cost comes from and how it can be distributed. The outermost loop is the sum over $n$: each term contributes to every entry of the covariance matrix, but terms with different $n$ are independent of one another, so we distribute them among parallel workers and add their contributions at the end. The cost of a single term grows with $n$, and for $n$ close to $n_{\rm max}$ each worker reaches a peak memory usage of 8--10\,GB. For $\ell_{\rm max}=30$, $p=25$ and $\Omega_{\rm K}=-10^{-3}$, computing a single full correlation matrix for an inhomogeneous lens space with an off-axis observer takes up to 4 hours of wall-clock time with 40 parallel workers on one node with two 2.2\,GHz AMD EPYC 7552 processors (96 physical cores, 192 hardware threads), with a total peak memory usage of about 350\,GB. The number of workers therefore sets a trade-off between memory and runtime: halving it halves the peak memory usage but roughly doubles the wall-clock time.

Empirically, the computation time grows as a power of $n_{\rm max}$ with exponent between 3 and 3.4, close to the $\mathcal{O}(n_{\rm max}^3)$ growth in the number of $(n,m_L,m_R)$ terms per matrix entry in \cref{eqn:clmlpmp_lens}. For observers on either symmetry circle, $\chi_0\in\{0,\pi/2\}$ (sufficient for homogeneous lens spaces, where all observers are equivalent), the Wigner matrices are (anti)diagonal, which means that the sum over $m_R$ collapses and the cost drops to $\mathcal{O}(n_{\rm max}^2)$. In the general case, since $n_{\rm max}\propto|\Omega_{\rm K}|^{-1/2}$ at fixed $\ell_{\rm max}$, the cost grows as $\sim|\Omega_{\rm K}|^{-1.5}$ towards the small curvatures favoured by current data.

\begin{figure}[htbp]
    \centering
    \includegraphics[width=0.85\linewidth]{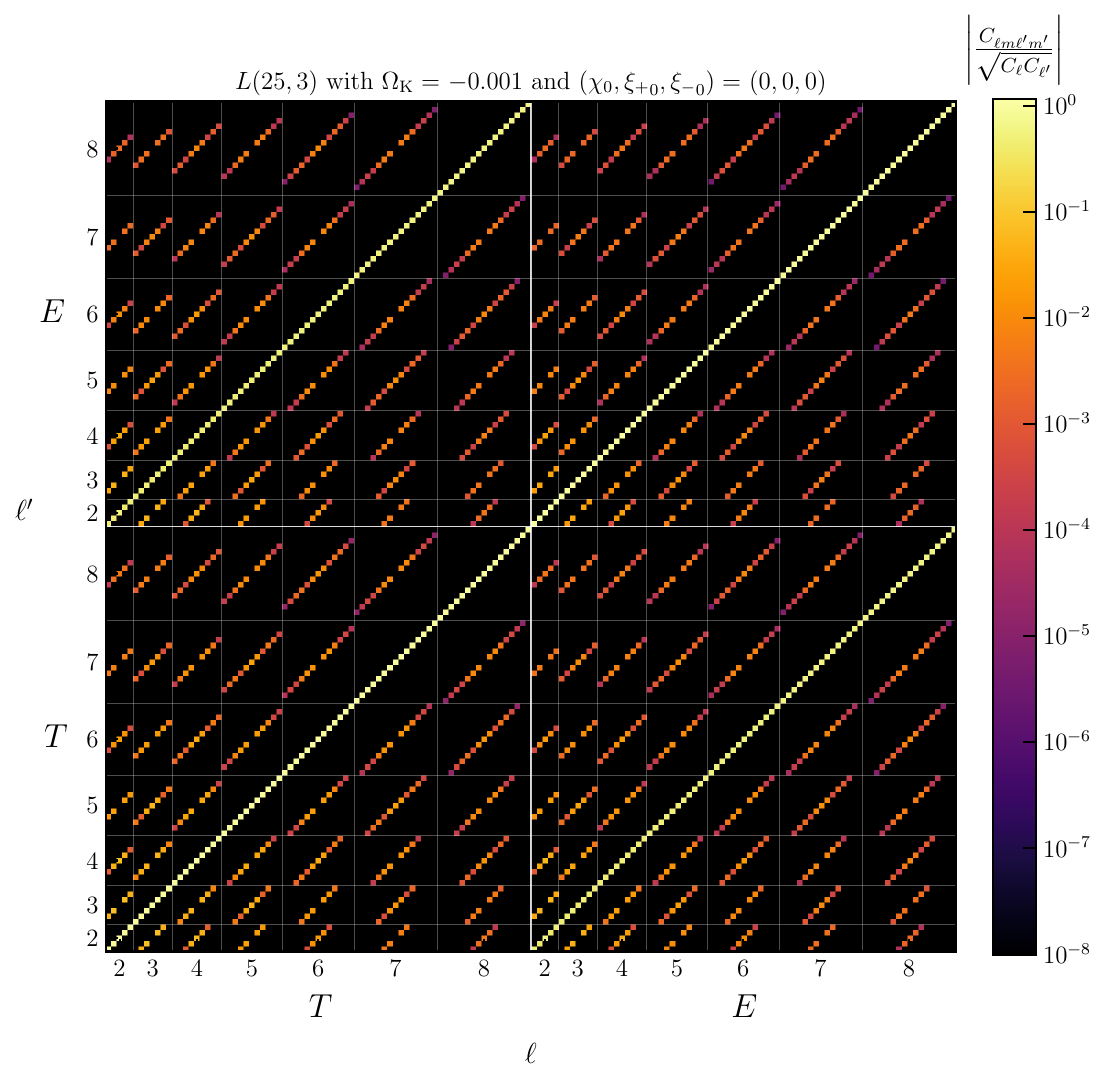}
    \caption{Modulus of the CMB covariance for an observer sitting at $(\chi_0,{\xi_+}_0,{\xi_-}_0)=(0,0,0)$ in $L(25,3)$ with $\Omega_{\rm K}=-10^{-3}$, normalized to the diagonal values of a three-sphere with the same curvature. The $TT$, $TE$ and $EE$ blocks are shown, ordered as indicated by the axis labels.}
    \label{fig:clmlpmp_onaxis}
\end{figure}

\begin{figure}[htbp]
    \centering
    \includegraphics[width=0.85\linewidth]{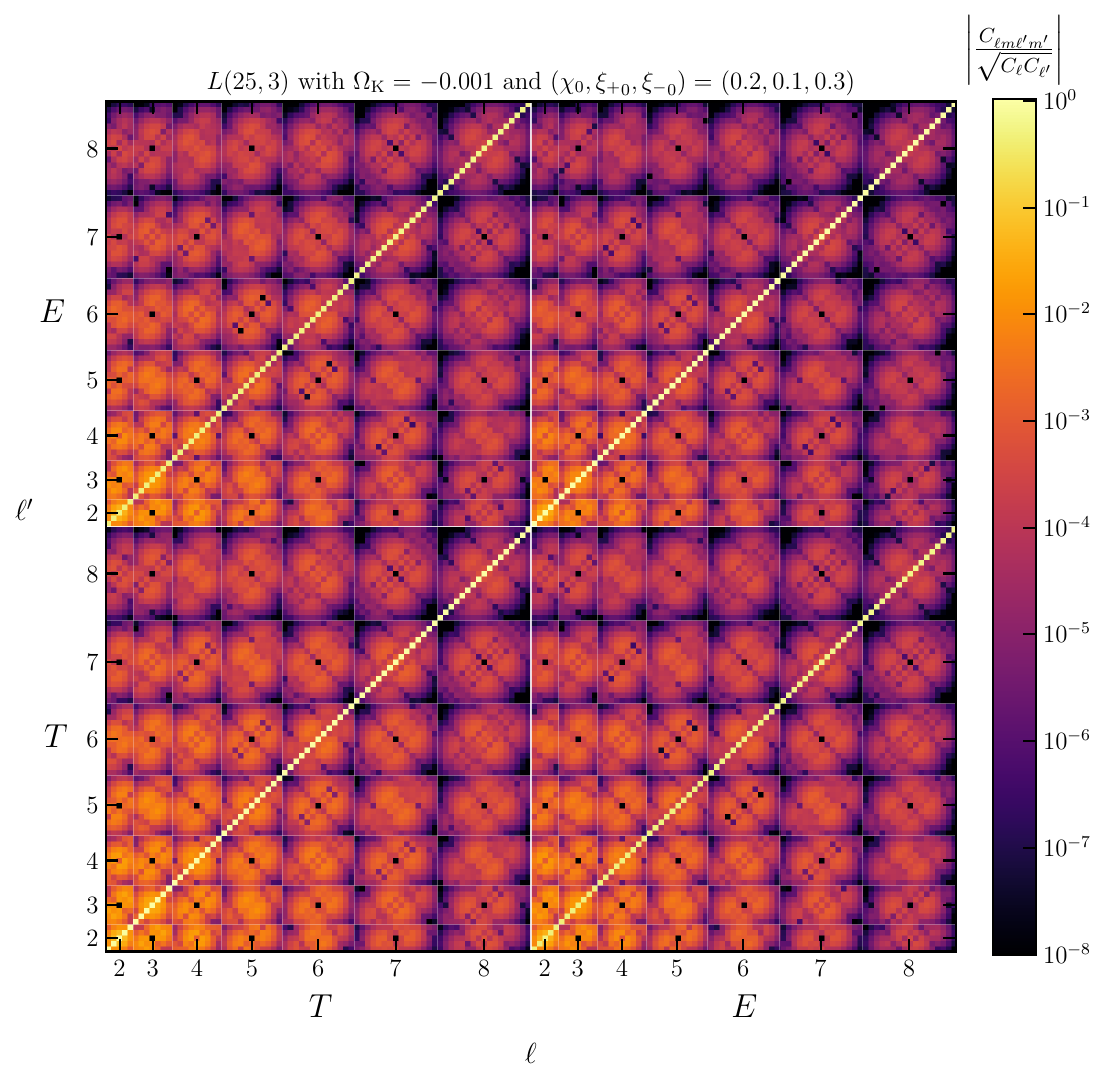}
    \caption{Modulus of the CMB covariance for an observer sitting at $(\chi_0,{\xi_+}_0,{\xi_-}_0)=(0.2,0.1,0.3)$ in $L(25,3)$ with $\Omega_{\rm K}=-10^{-3}$, with the same blocks and normalization as in \Cref{fig:clmlpmp_onaxis}.}
    \label{fig:clmlpmp_offaxis}
\end{figure}

Figures~\ref{fig:clmlpmp_onaxis} and \ref{fig:clmlpmp_offaxis} show representative temperature--polarization covariance matrices for the lens space $L(25,3)$ with $\Omega_{\rm K}=-10^{-3}$. \Cref{fig:clmlpmp_onaxis} corresponds to an observer located at one of the symmetry axes ($\chi_0=0$), for which the covariance matrix takes the simplified form discussed in \cref{ssec:corr_mats}. \Cref{fig:clmlpmp_offaxis} shows an off-axis observer, illustrating how changing the observer position mixes the toroidal eigenmodes through the Wigner D-matrices and produces a substantially richer correlation pattern. In both cases, the covariance matrix exhibits the characteristic off-diagonal correlations induced by the non-trivial spatial topology. Note that the on-axis matrix of \Cref{fig:clmlpmp_onaxis} is non-zero only for $m=m'$, since $\chi_0=0$ forces such a selection rule; the off-axis matrix of \Cref{fig:clmlpmp_offaxis} does not have such a restriction.

\subsection{Kullback--Leibler divergence}
\label{ssec:kl}

Before eventually confronting lens-space models with CMB observations, it is useful to quantify how distinguishable their predictions are from those of the fiducial simply connected model. If two cosmological models produce nearly identical probability distributions for the observables, no statistical analysis can reliably discriminate between them, regardless of the quality of the data.
A widely used measure of distinguishability between probability distributions is the KL divergence. For two distributions $P_1$ and $P_2$ of a random vector $\bm{x}$ it is defined as
\begin{equation}
    D_{\mathrm{KL}}(P_1 || P_2) = \int \mathrm{d}\bm{x} \,\, P_1(\bm{x}) \ln\, \left[\frac{P_1(\bm{x})}{P_2(\bm{x})} \right] .
\end{equation}
The KL divergence is always non-negative and vanishes if and only if the two distributions are identical almost everywhere. It is important to emphasize, however, that it is not a metric, since in general
$D_{\rm KL}(P_1||P_2)\neq D_{\rm KL}(P_2||P_1)$.

For the zero-mean real multivariate Gaussian distributions relevant to CMB analyses, the expression simplifies considerably. Writing the covariance matrices as $\mathbf C_{P_1}$ and $\mathbf C_{P_2}$, one obtains

\begin{equation}
 D_{\mathrm{KL}}(P_1 || P_2)=\frac{1}{2}\sum_i (\lambda_i-\log\lambda_i-1)\,, \quad \lambda_i\in\rm{Spec}(\mathbf C_{P_1} \mathbf C_{P_2}^{-1})\,.
\end{equation}

The KL divergence $D_{\rm KL}(P_1||P_2)$ gives the expected log-likelihood ratio in favour of $P_1$ when the data $\bm{x}$ are drawn from that model. A larger KL divergence indicates greater information, on average, for distinguishing a non-trivial topology from the covering space, but does not by itself guarantee a higher detection probability at a fixed false-positive rate. In particular, there is no universal KL divergence threshold for detectability, since the detection probability depends on the likelihood-ratio distributions as well as on the adopted search and detection statistics \cite{COMPACT:2026nfm}. Here, we adopt a KL divergence of order unity, measured in nats, as a loose guideline rather than an operational detection threshold. Exceeding unity therefore does not, by itself, establish that a manifold can be reliably detected.

\begin{figure}[htbp]
    \centering
    \begin{minipage}[t]{0.48\textwidth}
        \centering
        \includegraphics[width=\linewidth]{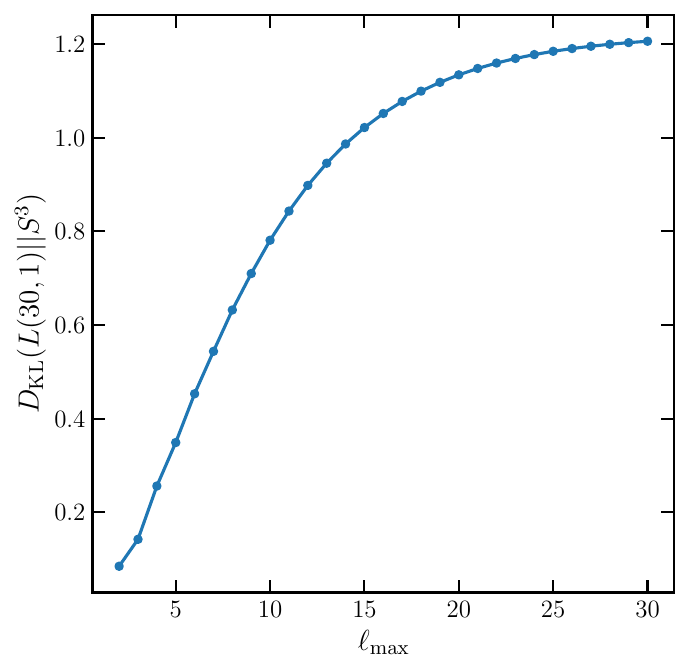}
    \end{minipage}\hfill
    \begin{minipage}[t]{0.487\textwidth}
        \centering
        \includegraphics[width=\linewidth]{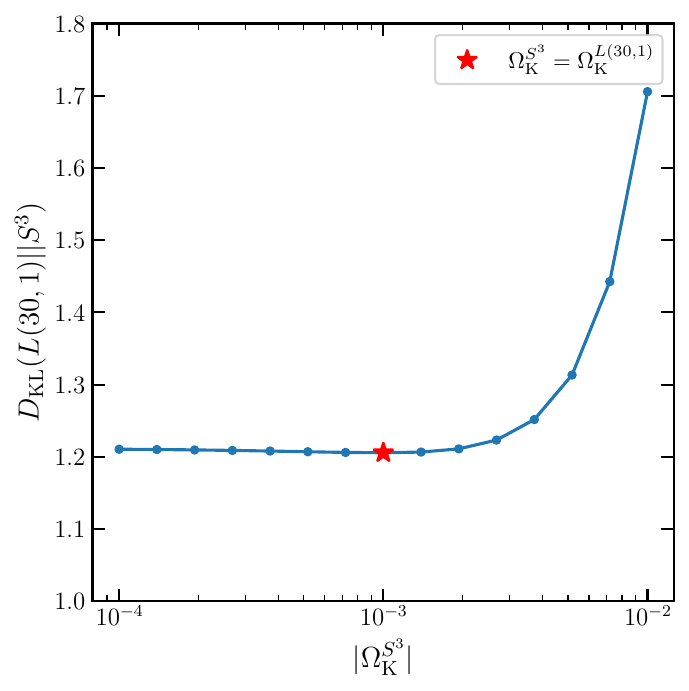}
    \end{minipage}
    \caption{\textit{Left:} KL divergence between a homogeneous lens space $L(30,1)$ and a three-sphere as a function of the maximum multipole $\ell$ included in the covariance matrices. The curvature density parameter used to produce this plot is $\Omega_{\rm K}=-10^{-3}$. \textit{Right:} KL divergence between a homogeneous lens space $L(30,1)$ and three-sphere as a function of the curvature parameter of the three-sphere. The red marker shows the value of $\Omega_{\rm K}$ used for the lens space.}
    \label{fig:kl_combined}
\end{figure}

The left panel of \Cref{fig:kl_combined} shows the KL divergence between the homogeneous lens space $L(30,1)$ and the simply connected three-sphere as a function of the maximum multipole included in the covariance matrices. As expected, distinguishability increases monotonically with $\ell_{\rm max}$ at first, reflecting the additional topological information encoded in progressively smaller angular scales. However, the information content saturates relatively quickly and ceases to increase by $\ell_{\rm max}\sim 30$, matching the behaviour found in Ref.~\cite{COMPACT:2023rkp} for compact Euclidean manifolds. Therefore, we only compute the correlation matrices up to $\ell_{\rm max}=30$. We stress that this saturation has been demonstrated for $L(30,1)$ at $\Omega_{\rm K}=-10^{-3}$, which has $d_{\rm NC}/d_{\rm LSS}>1$. For models with $d_{\rm NC}$ smaller than $d_{\rm LSS}$, the information does not saturate by $\ell_{\rm max}=30$, and the corresponding divergences reported below should be read as lower bounds.

Unlike in the Euclidean case, where the infinite covering space has no free topological parameters, in the spherical case, one can in principle ask how different a lens space with a certain curvature is from a three-sphere universe with a \textit{different} curvature parameter. To isolate the topological signal, by default we compute all KL divergences using the same curvature density parameter $\Omega_{\rm K}$ for the spaces being compared. However, we quantify the influence of curvature mismatch by studying the dependence of the KL divergence on spatial curvature of the three-sphere in the right panel of \Cref{fig:kl_combined}. The KL divergence is evaluated between a fixed lens-space model and simply connected manifolds with varying curvature parameter. The divergence reaches its minimum at the curvature of the lens-space model itself, where the difference between the distributions is purely topological. Changing the curvature of the three-sphere has a minor effect on $D_{\rm KL}$, unless the curvature mismatch becomes sufficiently large, in which case the geometrical difference between the models dominates the divergence. This motivates our choice to fix $\Omega_{\rm K}$ to the same value for the spaces being compared when assessing the topological contribution to the KL divergence.

\begin{figure}[htbp]
    \centering
    \includegraphics[width=\linewidth]{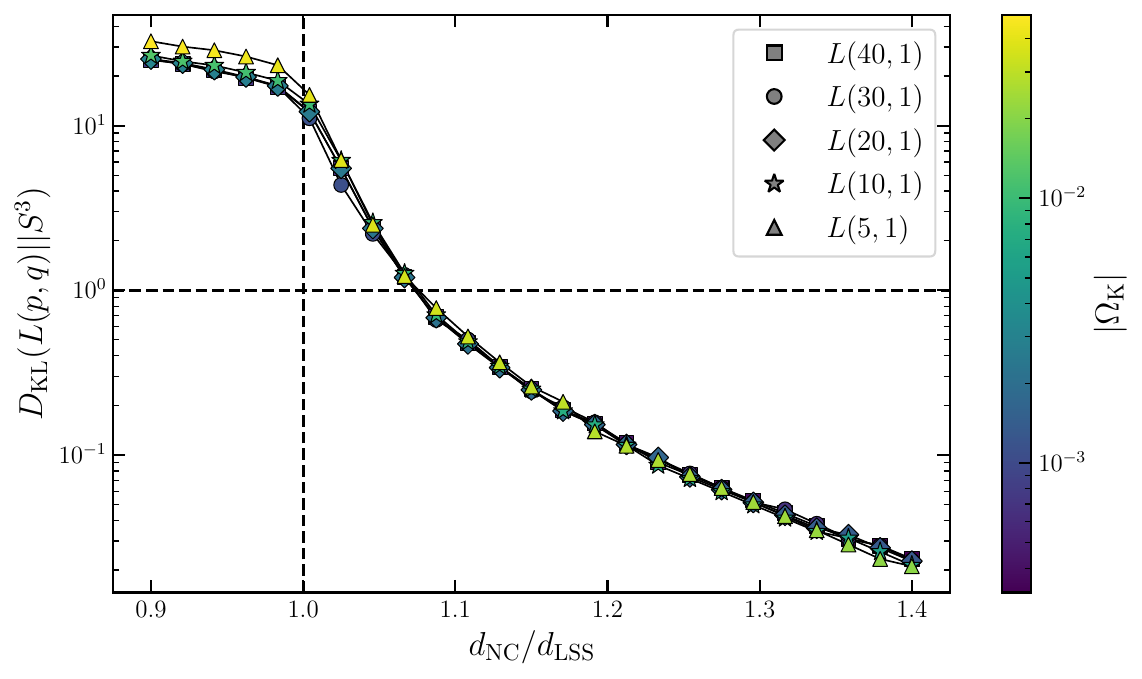}
    \caption{KL divergence between the homogeneous lens spaces $L(p,1)$, with $p=5,10,20,30,40$ indicated by the shapes of the markers, and the three-sphere with the same curvature, as a function of the nearest-clone distance relative to the diameter of the last-scattering surface, $d_{\rm NC}/d_{\rm LSS}$. At fixed $p$, each value of $d_{\rm NC}/d_{\rm LSS}$ corresponds to a different curvature $|\Omega_{\rm K}|$, indicated by the color of the markers. The vertical dashed line marks $d_{\rm NC}=d_{\rm LSS}$, below which matched circles appear in the sky. The horizontal dashed line marks $D_{\rm KL}=1$ for visual reference.}
    \label{fig:kl_homogeneous}
\end{figure}

\begin{figure}[htbp]
    \centering
    \includegraphics[width=\linewidth]{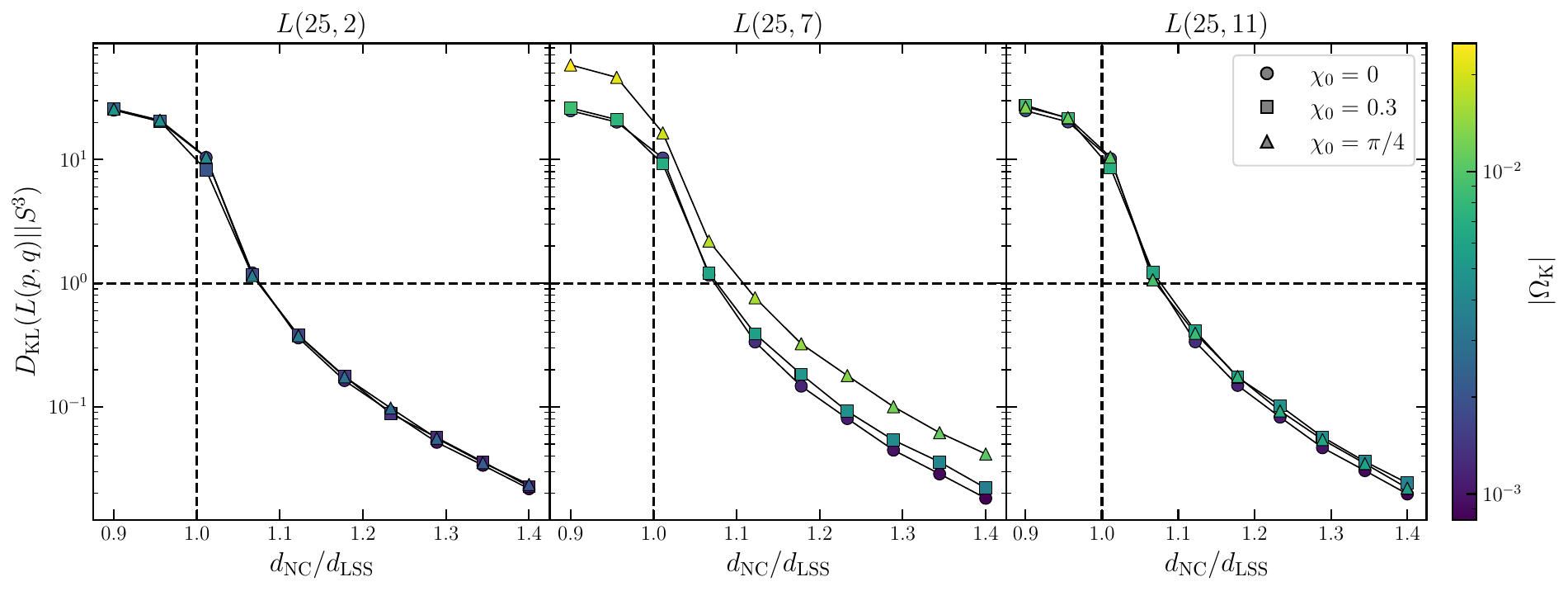}
    \caption{Same as Figure~\ref{fig:kl_homogeneous} for the inhomogeneous lens spaces $L(25,q)$ with $q=2,7,11$ (from left to right). The shapes of the markers denote the observer position $\chi_0$, and $d_{\rm NC}=d_{\rm NC}(\chi_0)$ is the nearest-clone distance of each observer. The results do not depend on ${\xi_+}_0$ and ${\xi_-}_0$, which we set to zero.}
    \label{fig:kl_inhomogeneous}
\end{figure}

Figures~\ref{fig:kl_homogeneous} and \ref{fig:kl_inhomogeneous} show the KL divergence between lens spaces and the three-sphere with the same curvature as a function of $d_{\rm NC}/d_{\rm LSS}$, for homogeneous and inhomogeneous lens spaces, respectively. For a given lens space and observer position, $d_{\rm NC}$ is a fixed fraction of the curvature radius, so each point in these figures corresponds to a different value of $|\Omega_{\rm K}|$, indicated by the color of the markers. The main conclusion of this section is that $d_{\rm NC}/d_{\rm LSS}$ is a nearly universal proxy for the amount of topological information retained in the CMB, and the divergence seems to depend on curvature, particular lens space and observer location almost exclusively through this ratio. As mentioned before, in the regime $d_{\rm NC}<d_{\rm LSS}$, where matched circles are present in the sky, the divergence is large but has not saturated at $\ell_{\rm max}=30$, so the trend in that part of the figures is sensitive to the multipole cut. We therefore refrain from reading off absolute information contents there.

For the homogeneous lens spaces $L(p,1)$ with $p\in\{5,10,20,30,40\}$, shown in Figure~\ref{fig:kl_homogeneous}, the curves for different $p$ are almost indistinguishable for $d_{\rm NC}\gtrsim d_{\rm LSS}$. All of them cross $D_{\rm KL}=1$ at $d_{\rm NC}/d_{\rm LSS}\simeq1.07$, even though the corresponding curvature ranges from $|\Omega_{\rm K}|\simeq6\times10^{-4}$ for $L(40,1)$ to $|\Omega_{\rm K}|\simeq3\times10^{-2}$ for $L(5,1)$. Beyond this point, the divergence falls steeply by about an order of magnitude for every increase of $0.15$ in $d_{\rm NC}/d_{\rm LSS}$, down to $D_{\rm KL}\simeq0.02$ at $d_{\rm NC}/d_{\rm LSS}=1.4$. This universality across $p$ is not surprising. In a homogeneous lens space, every observer sees the same pattern of clones: they come in pairs along a single great circle through the observer, at distances $j\cdot d_{\rm NC}$ with $j=1,\ldots,\lfloor p/2\rfloor$, and the $j$-th pair is rotated by an angle $\pm2\pi j/p$ about that circle. At fixed $d_{\rm NC}/d_{\rm LSS}$, increasing $p$ only increases the curvature radius, $R_c=p\,d_{\rm NC}/2\pi$, and reduces the rotation of the clones. The fact that even $L(5,1)$, with $|\Omega_{\rm K}|\simeq0.02$--$0.05$ and its nearest clones rotated by $72^\circ$, follows the same curve shows that neither the background curvature nor the rotation of the clones has an appreciable impact on $D_{\rm KL}$ at fixed $d_{\rm NC}/d_{\rm LSS}$. This is in line with the insensitivity to such rotations found for on-axis observers in compact Euclidean manifolds~\cite{COMPACT:2023rkp}. 

For the inhomogeneous lens spaces $L(25,q)$ with $q\in\{2,7,11\}$, shown in Figure~\ref{fig:kl_inhomogeneous}, we compute the divergence for observers at $\chi_0\in\{0,0.3,\pi/4\}$. It is not necessary to vary ${\xi_+}_0$ and ${\xi_-}_0$, as the rotations $(\xi_+,\xi_-)\to(\xi_++\Delta\xi_+,\xi_-+\Delta\xi_-)$ commute with the generator of \cref{eqn:lens_isom} and thus descend to isometries of $L(p,q)$, so that all observers with the same $\chi_0$ are equivalent up to a rotation of their reference frame, under which $D_{\rm KL}$ is invariant (see \cref{ssec:location}).

The observer position affects the divergence in two distinct ways. First, it changes $d_{\rm NC}$ itself. Since the right-hand side of \cref{eqn:lens_clone_distance} is a weighted average of $\cos(2\pi j/p)$ and $\cos(2\pi jq/p)$, every observer has $d_{\rm NC}(\chi_0)\geq2\pi R_c/p$, with equality only for observers on the axes, $\chi_0\in\{0,\pi/2\}$. Off-axis observers can have considerably more distant clones. For instance, the observer at $\chi_0=\pi/4$ in $L(25,7)$ has $d_{\rm NC}\simeq3.5\times2\pi R_c/25$. At fixed curvature, off-axis observers are therefore less sensitive to the topology, and reaching a given value of $d_{\rm NC}/d_{\rm LSS}$ requires a larger $|\Omega_{\rm K}|$, as reflected by the colors in Figure~\ref{fig:kl_inhomogeneous}. This effect is absorbed by plotting the divergence against $d_{\rm NC}/d_{\rm LSS}$.

Second, at fixed $d_{\rm NC}/d_{\rm LSS}$, the divergence depends on how many clones lie at distances comparable to $d_{\rm NC}$. On-axis observers have exactly the same clone positions as observers in a homogeneous lens space with the same $p$, the only difference being the rotation of the $j$-th clone, now by $2\pi jq/p$. Accordingly, the three on-axis curves are practically indistinguishable from one another and from the homogeneous curves of Figure~\ref{fig:kl_homogeneous}, crossing $D_{\rm KL}=1$ at $d_{\rm NC}/d_{\rm LSS}\simeq1.06$. Away from the axes, the clones of the observer spread over the torus $\chi=\chi_0$, on which they form a two-dimensional lattice, and additional clones appear at distances comparable to $d_{\rm NC}$. Changing $q$, for a fixed $p$, has the effect of modifying this lattice. For $L(25,2)$, this lattice is very elongated: the clones remain almost aligned, with the second pair at $\simeq2\,d_{\rm NC}$, and the three observers give nearly identical results. For $L(25,11)$, off-axis observers have a second pair of clones at $\simeq1.2\,d_{\rm NC}$ ($\chi_0=0.3$) or three pairs at $\simeq2\,d_{\rm NC}$ ($\chi_0=\pi/4$), but the divergence hardly varies. The most interesting case is $L(25,7)$ at $\chi_0=\pi/4$. Since $7^2\equiv-1\pmod{25}$, the orbit of this observer on the Clifford torus is invariant under a quarter turn and forms a square lattice, so that the observer has two pairs of equidistant nearest clones in independent directions. This configuration yields the largest divergence in our sample (triangle markers in the central panel of Figure~\ref{fig:kl_inhomogeneous}), about a factor of two larger than the rest of the cases, although $D_{\rm KL}$ drops below unity at $d_{\rm NC}/d_{\rm LSS}\simeq1.11$. The enhancement of the information content for observers with several equidistant nearest clones mirrors what was found for off-axis observers in Euclidean manifolds~\cite{COMPACT:2023rkp}.

We emphasize that all the clones of an observer at $\chi_0$ lie on the torus $\chi=\chi_0$, which is intrinsically flat and on which the generator of \cref{eqn:lens_isom} acts as a translation. In the limit $R_c\to\infty$ at fixed $d_{\rm NC}/d_{\rm LSS}$, the clone pattern within the observable Universe therefore approaches that of a non-compact Euclidean manifold. When the nearby clones are aligned, as in homogeneous lens spaces and for on-axis observers, this is a slab space, \slabh, or its twisted version, \slabi, with a rotation between successive clones set by $q$. When they form a two-dimensional lattice, as for off-axis observers in inhomogeneous lens spaces, it is a chimney space, \E{11}, whose cross-section is set by $q$ and $\chi_0$. The square lattice of $L(25,7)$ at $\chi_0=\pi/4$, for instance, is the analogue of a chimney space with a square cross-section. Since the clones never leave this two-dimensional torus, the clone pattern of a lens space can never resemble that of a compact Euclidean manifold, whose clones span three independent directions. The near-independence of $D_{\rm KL}$ from $p$ at fixed $d_{\rm NC}/d_{\rm LSS}$ in \cref{fig:kl_homogeneous}, over curvatures spanning almost two orders of magnitude, suggests that the background curvature is already negligible for the information content at the values considered here. A direct comparison of the covariance matrices with those of the corresponding Euclidean manifolds is left for future work.

Finally, we stress the assumptions underlying all the divergences in this section. The lens-space and the three-sphere share the same primordial spectrum $P^{\mathcal{R}}_{S^3}(k_n)$, and in both cases the amplitudes of the allowed eigenmodes are independent, with a variance that depends only on the eigenvalue label $n$ and not on the remaining labels. This is the minimal choice: the two models differ only through the selection of eigenmodes in \cref{eqn:selection_rule_lens}, so that $D_{\rm KL}$ quantifies the information carried by the topology alone, without any additional assumption about the physics of the early Universe. Mechanisms that make the primordial statistics themselves depend on the topology, such as a modified spectrum on scales comparable to $d_{\rm NC}$ or other imprints of the topology on the early Universe~\cite{COMPACT:2026pqs,COMPACT:2026vdj,Negro:2026ooh}, would change these results, but not necessarily by increasing the divergence. In compact Euclidean manifolds, departures from a power-law spectrum can either enhance the topological imprints in the CMB or suppress them~\cite{COMPACT:2026pqs}. Our results should therefore be read as the detectability of lens-space topology under the most economical assumptions about the primordial perturbations, not as a strict bound in either direction.

\section{Conclusions}
\label{sec:conclusions}

In this work, we have developed the necessary formalism and machinery to compute harmonic two-point functions of cosmological observables in  lens spaces $L(p,q)$ of arbitrary parameters and arbitrary observer location. We compute for the first time all the off-diagonal entries of the CMB covariance matrices $C_{\lmlpmp}^{XY;pq}$.

We have computed the Kullback-Leibler (KL) divergence between the Gaussian distributions of CMB harmonic coefficients predicted in a lens space and in a simply connected three-sphere with the same curvature, which can be read as the expected log-likelihood ratio in favour of the lens space when it is the true model. We evaluated it in an idealized setting, with full-sky, noiseless, unlensed, cosmic-variance-limited temperature and $E$-mode data up to $\ell_{\rm max}=30$. We find that topological signatures can remain significant, with $D_{\rm KL}>1$, even when the nearest-clone distance $d_{\rm NC}$ exceeds the diameter of the last-scattering surface, $d_{\rm LSS}$, although the divergence falls off rapidly for $d_{\rm NC}/d_{\rm LSS}\gtrsim1.1$. For homogeneous lens spaces, the KL divergence at fixed $d_{\rm NC}/d_{\rm LSS}$ is nearly independent of $p$, and hence of $\Omega_{\rm K}$, whereas for inhomogeneous lens spaces it also depends on the observer's position. 
While a direct comparison is not trivial, the behaviour found here is quantitatively similar to that reported in our previous investigations of Euclidean manifolds~\cite{COMPACT:2022gbl, COMPACT:2023rkp, COMPACT:2026nez}, where the
divergence likewise drops below unity once $d_{\rm NC}$ exceeds $d_{\rm LSS}$ by a similar margin.

A notable property of the lens spaces collectively is that curvature alone does not set the topology scale. For any value of the curvature, however small, there are lens spaces of arbitrarily small volume in which every observer's nearest clone is also arbitrarily close. This is not surprising, given that the same holds true in Euclidean manifolds, which have zero spatial curvature. Since $D_{\rm KL}$ depends essentially on $d_{\rm NC}/d_{\rm LSS}$, such nearly flat lens spaces would be as detectable as their more curved counterparts. As the curvature vanishes, the clone pattern of a lens space within the observable Universe approaches that of a slab or chimney Euclidean space, but never that of a compact one (see \cref{ssec:kl}). Confirming this correspondence at the level of the covariance matrices is an interesting direction for future work.

The framework outlined in this paper offers several applications to the study of cosmic topology. 
The framework can be implemented into future Bayesian analyses, either to constrain cosmic topology with already existing CMB datasets, such as the \textit{Planck} temperature and polarization maps, or to perform forecast analyses for upcoming experiments such as LiteBIRD. 
However, computing one full correlation matrix for an inhomogeneous lens space with an off-axis observer currently takes a few hours on a full 96-core node, with a few hundred GB of memory, and the cost grows steeply with $n_{\rm max}$, and hence towards small $|\Omega_{\rm K}|$ (see \cref{ssec:eval_cov_mat}). A search on real data would require such matrices over a grid in $(p,q,\chi_0,\Omega_{\rm K})$, so computational cost is the main practical obstacle. Machine learning methods may therefore be useful to increase computation speed, as explored in Ref.~\cite{COMPACT:2024dqe}. 

Furthermore, the theoretical calculations may be adapted to other classes of 3D spherical manifolds, in particular, the prism spaces or the binary polyhedral spaces of the three-sphere. Ultimately, the results presented in this work are a necessary step towards a comprehensive likelihood search for the topology of our Universe.

\acknowledgments

We acknowledge the use of large language models (LLMs), including ChatGPT and Claude, to assist with the design, optimization and implementation of parts of the computational code used in this work. All AI-assisted code was reviewed, tested, and validated by the authors, who take full responsibility for the correctness of the implementation and the results presented in this paper.

This work is partially funded by the European Commission - NextGenerationEU, through Momentum CSIC Programme: Develop Your Digital Talent. Numerical calculations have been performed on the Faculty of Arts and Sciences Research Computing (FASRC) Cannon Cluster at Harvard University and the Hydra Cluster at IFT. We acknowledge HPC support by Emilio Ambite, staff hired under the Generation D initiative, promoted by Red.es, an organisation attached to the Spanish Ministry for Digital Transformation and the Civil Service, for the attraction and retention of talent through grants and training contracts, financed by the Recovery, Transformation and Resilience Plan through the European Union's Next Generation funds.
M.M.B. acknowledges support by the Spanish Ministry of Science, Innovation and Universities under the FPU predoctoral grant FPU22/02306.
A.N.\ is supported by the Richard S.\ Morrison Fellowship.  
G.D.S. was supported by DOE award DESC0009946.
Y.A. acknowledges support by the Spanish Research Agency (Agencia Estatal de Investigaci\'on)'s grant RYC2020-030193-I/AEI/10.13039/501100011033, by the European Social Fund (Fondo Social Europeo) through the  Ram\'{o}n y Cajal program within the State Plan for Scientific and Technical Research and Innovation (Plan Estatal de Investigaci\'on Cient\'ifica y T\'ecnica y de Innovaci\'on) 2017-2020, by the Spanish Research Agency through the grant IFT Centro de Excelencia Severo Ochoa No CEX2020-001007-S funded by MCIN/AEI/10.13039/501100011033, by the Spanish National Research Council (CSIC) through the Talent Attraction grant 20225AT025, and by the Spanish Research Agency's Consolidaci\'on Investigadora
2024 grant CNS2024-154430.
J.C.D. is supported by the Spanish Research Agency (Agencia Estatal de Investigaci\'on), the Ministerio de Ciencia, Inovaci\'on y Universidades, and the European Social Funds through grant JDC2023-052152-I, as part of the Juan de la Cierva program.
G.A. is supported by the Spanish Research Agency's Consolidaci\'on Investigadora
2024 grant CNS2024-154430.
F.C.G\ was supported by the Presidential Society of STEM Postdoctoral Fellowship at Case Western Reserve University and by Ministerio de Ciencia, Innovaci\'on y Universidades, Spain, through a Beatriz Galindo Junior grant BG23/00061.
D.P.M. acknowledges support by the Bulgarian National Science Fund program ``VIHREN--2024" project No. KP--06--DV/9/17.12.2024.
J.J.O.G acknowledges support by the Spanish Ministry of Science, Innovation and Universities under the FPU predoctoral grant FPU24/02041.
A.T. is supported by the European Union's Horizon Europe research and innovation programme under the Marie Sk\l odowska-Curie grant agreement No. 101126636.
J.C.D. is and A.T. was supported by CSIC through grant No. 20225AT025.
G.D.S., C.J.C., A.K. and D.P.M.\ acknowledge partial support from NASA ATP grant 80NSSC22K0805 and from NASA ADAP grant 24-ADAP24-0018;  A.H.J.\ from the Royal Society and STFC in the UK; and A.S. from DOE grant DESC0009946.
MMB and CP thank CWRU and
GDS thanks the IFT for their hospitality while working on this project.
\begin{appendices}

\renewcommand{\appendix}{\par
    \setcounter{section}{0}
    \setcounter{subsection}{0}
    \gdef\thesection{\Alph{section}}
}

\section{Quaternions}
\label{app:quaternions}

Many of the geometric properties of the three-sphere become particularly transparent when the three-sphere is identified with the Lie group of unit quaternions. In this appendix we summarize the quaternionic identities used throughout the paper and derive the decomposition of the isometry group of the three-sphere.\\

The quaternion algebra is
\begin{equation}
\mathbb{H} = \left\{ a+b\mathbf{i}+c\mathbf{j}+d\mathbf{k}\,\middle|\,a,b,c,d\in\mathbb{R},\;\mathbf{i}^2=\mathbf{j}^2=\mathbf{k}^2=\mathbf{ijk}=-1\right\}\,.
\end{equation}
The subset of unit quaternions,
\begin{equation}
\mathbb{H}_1=\{\mathbf{g}\in\mathbb{H}:\|\mathbf{g}\|=1\}\,,
\end{equation}
forms a Lie group isomorphic to $\mathrm{SU}(2)$ and is naturally identified with the unit three-sphere. Writing a quaternion as
\begin{equation}
\mathbf{g}=(g,\vec g)=g+\vec g\,,
\end{equation}
with scalar part $g$ and vector part $\vec g\in\mathbb{R}^3$, the Hamilton product takes the compact form
\begin{equation}
\mathbf{g}\mathbf{h}=\left(gh-\vec g\,\cdot\,\vec h,\,g\vec h+h\vec g+\vec g\times\vec h\right)\,,
\end{equation}
from which the non-commutativity follows directly,
\begin{equation}
[\mathbf g,\mathbf h]=2\,\vec g\times\vec h\,.
\end{equation}
The conjugate of a quaternion is
\begin{equation}
\mathbf g^*=(g,-\vec g)\,,
\end{equation}
and for unit quaternions
\begin{equation}
\mathbf g^{-1}=\mathbf g^*\,.
\end{equation}
The geodesic distance on the three-sphere is
\begin{equation}
d(\mathbf g,\mathbf h)=\arccos\,\left[(\mathbf g\mathbf h^*)_s\right]\,,
\end{equation}
where $(\cdot)_s$ denotes the scalar part. Therefore,
\begin{equation}
d(\mathbf g,\mathbf u\mathbf g)=\arccos\,\left[(\mathbf u)_s\right]\,,
\end{equation}
which is independent of $\mathbf g$. Left (and similarly right) multiplication by a unit quaternion therefore defines a Clifford translation of the three-sphere. Every unit quaternion admits the exponential representation
\begin{equation}
\mathbf u=\exp(a\mathbf r)=\cos a+\sin a\,\mathbf r\,,
\end{equation}
where $\mathbf r$ is a unit purely imaginary quaternion. Conjugation,
\begin{equation}
\mathbf g\longmapsto\mathbf u^{-1}\mathbf g\mathbf u\,,
\end{equation}
acts on the imaginary part of $\mathbf g$ as a three-dimensional rotation through angle $2a$ about the axis $\mathbf r$. Indeed,
\begin{equation}
\mathbf u^{-1}\mathbf g\mathbf u=\mathbf g\iff[\mathbf g,\mathbf u]=0\iff\vec g\times\vec r=0\,,
\end{equation}
so the fixed vectors are precisely those parallel to the rotation axis.

\paragraph{Isometries of the three-sphere.}
Every orientation-preserving isometry of the three-sphere can be written uniquely up to a common sign as
\begin{equation}
\phi_{(\mathbf u,\mathbf v)}:\mathbf g\mapsto\mathbf u\mathbf g\mathbf v\,,
\end{equation}
with $\mathbf u,\mathbf v\in\mathbb H_1$. Since
\begin{equation}
(\mathbf u,\mathbf v)\sim(-\mathbf u,-\mathbf v)\,,
\end{equation}
one obtains the well-known group isomorphism
\begin{equation}
\mathrm{SO}(4)\cong(\mathrm{SU}(2)\times\mathrm{SU}(2))/\mathbb Z_2\cong(S^3\times S^3)/\mathbb Z_2\,.
\end{equation}
Left and right Clifford translations correspond to the special cases $\mathbf v=\mathbf1$ and $\mathbf u=\mathbf1$, respectively. Since the Wigner $D$-matrices furnish irreducible representations of $\mathrm{SU}(2)$, they satisfy
\begin{equation}
D^j(\mathbf u\mathbf g\mathbf v)=D^j(\mathbf u)\,D^j(\mathbf g)\,D^j(\mathbf v)\,.
\end{equation}
This representation property allows the action of any spherical manifold group $\Gamma$ to be implemented algebraically on the toroidal eigenmodes. In particular, the eigenmodes of a quotient manifold are obtained by restricting to the invariant subspace under the action of every $(\mathbf u,\mathbf v)\in\Gamma$, leading directly to the selection rules derived in Sec.~\ref{sec:eigenmodes}. When no individual basis vector is invariant, invariant eigenmodes are obtained by suitable linear combinations, exactly as plane waves are combined to construct eigenmodes of Euclidean manifolds with screw or corkscrew identifications.

\end{appendices}

\bibliographystyle{utphys}
\bibliography{topology,additional}

@article{Anselmi:2022uvj,
    author = "Anselmi, Stefano and Carney, Matthew F. and Giblin, John T. and Kumar, Saurabh and Mertens, James B. and O'Dwyer, Marcio and Starkman, Glenn D. and Tian, Chi",
    title = "{What is flat {\ensuremath{\Lambda}}CDM, and may we choose it?}",
    eprint = "2207.06547",
    archivePrefix = "arXiv",
    primaryClass = "astro-ph.CO",
    doi = "10.1088/1475-7516/2023/02/049",
    journal = "JCAP",
    volume = "02",
    pages = "049",
    year = "2023"
}

@ARTICLE{COMPACT:2022gbl,
        author = "Akrami, Yashar and others",
        collaboration = "COMPACT",
        title = "{Promise of Future Searches for Cosmic Topology}",
      journal = {Phys. Rev. Lett.},
         year = 2024,
        month = apr,
       volume = {132},
       number = {17},
          eid = {171501},
        pages = {171501},
          doi = {10.1103/PhysRevLett.132.171501},
archivePrefix = {arXiv},
       eprint = {2210.11426},
 primaryClass = {astro-ph.CO},
       adsurl = {https://ui.adsabs.harvard.edu/abs/2024PhRvL.132q1501A}
}

@article{lehoucq2004eigenmodeslensprismspaces,
  title={Eigenmodes of lens and prism spaces},
  author={Roland Lehoucq and Jean-Philippe Uzan and Jeffrey Weeks},
  journal={Kodai Mathematical Journal},
  volume={26},
  number={1},
  pages={119-136},
  year={2003},
  doi={10.2996/kmj/1050496653}
}

@article{Johansson:2015cca,
    author = "Johansson, H. T. and Forss{\'e}n, C.",
    title = "{Fast and Accurate Evaluation of Wigner 3$j$, 6$j$, and 9$j$ Symbols Using Prime Factorization and Multiword Integer Arithmetic}",
    eprint = "1504.08329",
    archivePrefix = "arXiv",
    primaryClass = "physics.comp-ph",
    doi = "10.1137/15M1021908",
    journal = "SIAM J. Sci. Comput.",
    volume = "38",
    number = "1",
    pages = "A376--A384",
    year = "2016"
}

@article{Lachieze-Rey:2005nyz,
    author = "Lachieze-Rey, Marc and Caillerie, S.",
    title = "{Laplacian eigenmodes for spherical spaces}",
    eprint = "astro-ph/0501419",
    archivePrefix = "arXiv",
    doi = "10.1088/0264-9381/22/4/004",
    journal = "Class. Quant. Grav.",
    volume = "22",
    pages = "695--708",
    year = "2005"
}

@article{Caillerie:2007gd,
    author = "Caillerie, S. and Lachieze-Rey, M. and Luminet, J. -P. and Lehoucq, R. and Riazuelo, A. and Weeks, J.",
    title = "{A new analysis of Poincar\'e dodecahedral space model}",
    eprint = "0705.0217",
    archivePrefix = "arXiv",
    primaryClass = "astro-ph",
    doi = "10.1051/0004-6361:20077656",
    journal = "Astron. Astrophys.",
    volume = "476",
    pages = "691",
    year = "2007"
}

@article{Aurich:2011wb,
    author = "Aurich, R. and Kramer, P. and Lustig, S.",
    title = "{CMB radiation in an inhomogeneous spherical space}",
    eprint = "1107.5214",
    archivePrefix = "arXiv",
    primaryClass = "astro-ph.CO",
    doi = "10.1088/0031-8949/84/05/055901",
    journal = "Phys. Scripta",
    volume = "84",
    pages = "055901",
    year = "2011"
}

@article{Aurich:2012sp,
    author = "Aurich, Ralf and Lustig, Sven",
    title = "{A survey of lens spaces and large scale CMB anisotropy}",
    eprint = "1203.4086",
    archivePrefix = "arXiv",
    primaryClass = "astro-ph.CO",
    doi = "10.1111/j.1365-2966.2012.21363.x",
    journal = "Mon. Not. Roy. Astron. Soc.",
    volume = "424",
    pages = "1556--1562",
    year = "2012"
}

@article{Aurich:2005hg,
    author = "Aurich, Ralf and Lustig, S. and Steiner, F.",
    title = "{The circles-in-the-sky signature for three spherical universes}",
    eprint = "astro-ph/0510847",
    archivePrefix = "arXiv",
    doi = "10.1111/j.1365-2966.2006.10296.x",
    journal = "Mon. Not. Roy. Astron. Soc.",
    volume = "369",
    pages = "240--248",
    year = "2006"
}

@article{Aurich:2004fq,
    author = "Aurich, Ralf and Lustig, Sven and Steiner, Frank",
    title = "{CMB anisotropy of the Poincare dodecahedron}",
    eprint = "astro-ph/0412569",
    archivePrefix = "arXiv",
    doi = "10.1088/0264-9381/22/11/010",
    journal = "Class. Quant. Grav.",
    volume = "22",
    pages = "2061--2083",
    year = "2005"
}

@article{Aurich:2012yd,
    author = "Aurich, Ralf and Lustig, Sven",
    title = "{Cosmic Topology of Prism Double-Action Manifolds}",
    eprint = "1205.0660",
    archivePrefix = "arXiv",
    primaryClass = "astro-ph.CO",
    doi = "10.1088/0264-9381/29/21/215005",
    journal = "Class. Quant. Grav.",
    volume = "29",
    pages = "215005",
    year = "2012"
}

@article{Uzan:2003ea,
    author = "Uzan, Jean-Philippe and Riazuelo, Alain and Lehoucq, Roland and Weeks, Jeffrey",
    title = "{Cosmic microwave background constraints on lens spaces}",
    eprint = "astro-ph/0303580",
    archivePrefix = "arXiv",
    doi = "10.1103/PhysRevD.69.043003",
    journal = "Phys. Rev. D",
    volume = "69",
    pages = "043003",
    year = "2004"
}

@article{Threlfall1931,
  author  = {Threlfall, W. and Seifert, H.},
  title   = {Topologische Untersuchung der Diskontinuit{\"a}tsbereiche endlicher Bewegungsgruppen des dreidimensionalen sph{\"a}rischen Raumes},
  journal = {Mathematische Annalen},
  year    = {1931},
  volume  = {104},
  pages   = {1--70},
  doi     = {10.1007/bf01457920}
}

@article{Gausmann:2001aa,
    author = "Gausmann, Evelise and Lehoucq, Roland and Luminet, Jean-Pierre and Uzan, Jean-Philippe and Weeks, Jeffrey",
    title = "{Topological lensing in spherical spaces}",
    eprint = "gr-qc/0106033",
    archivePrefix = "arXiv",
    doi = "10.1088/0264-9381/18/23/311",
    journal = "Class. Quant. Grav.",
    volume = "18",
    pages = "5155--5186",
    year = "2001"
}

@Inbook{Stanton1990,
author="Stanton, Dennis",
editor="Nevai, Paul",
title="An Introduction to Group Representations and Orthogonal Polynomials",
bookTitle="Orthogonal Polynomials: Theory and Practice",
year="1990",
publisher="Springer Netherlands",
address="Dordrecht",
pages="419--433",
isbn="978-94-009-0501-6",
doi="10.1007/978-94-009-0501-6_19",
url="https://doi.org/10.1007/978-94-009-0501-6_19"
}

@article{Kosowsky:1998nc,
    author = "Kosowsky, Arthur",
    title = "{Efficient computation of hyperspherical bessel functions}",
    eprint = "astro-ph/9805173",
    archivePrefix = "arXiv",
    reportNumber = "RAP-235",
    month = "5",
    year = "1998"
}

@article{Harrison:1967zza,
    author = "Harrison, E. R.",
    title = "{Normal Modes of Vibrations of the Universe}",
    doi = "10.1103/RevModPhys.39.862",
    journal = "Rev. Mod. Phys.",
    volume = "39",
    pages = "862--882",
    year = "1967"
}

@article{Lesgourgues:2013bra,
    author = "Lesgourgues, Julien and Tram, Thomas",
    title = "{Fast and accurate CMB computations in non-flat FLRW universes}",
    eprint = "1312.2697",
    archivePrefix = "arXiv",
    primaryClass = "astro-ph.CO",
    reportNumber = "CERN-PH-TH-2013-298, LAPTH-071-13",
    doi = "10.1088/1475-7516/2014/09/032",
    journal = "JCAP",
    volume = "09",
    pages = "032",
    year = "2014"
}

@article{Tram:2013xpa,
    author = "Tram, Thomas",
    title = "{Computation of hyperspherical Bessel functions}",
    eprint = "1311.0839",
    archivePrefix = "arXiv",
    primaryClass = "astro-ph.IM",
    doi = "10.4208/cicp.OA-2016-0071",
    journal = "Commun. Comput. Phys.",
    volume = "22",
    number = "3",
    pages = "852--862",
    year = "2017"
}

@book{Varshalovich:1988ifq,
    author = "Varshalovich, D. A. and Moskalev, A. N. and Khersonskii, V. K.",
    title = "{Quantum Theory of Angular Momentum}",
    doi = "10.1142/0270",
    publisher = "World Scientific Publishing Company",
    year = "1988"
}

@article{BALCH2021101712,
title = {Distributions of distances and volumes of balls in homogeneous lens spaces},
journal = {Differential Geometry and its Applications},
volume = {74},
pages = {101712},
year = {2021},
issn = {0926-2245},
doi = {https://doi.org/10.1016/j.difgeo.2020.101712},
url = {https://www.sciencedirect.com/science/article/pii/S0926224520301212},
author = {Brenden Balch and Chris Peterson and Clayton Shonkwiler},
}

@article{COMPACT:2025adc,
    author = "Copi, Craig J. and others",
    collaboration = "COMPACT",
    title = "{Cosmic topology. Part IIb. Eigenmodes, correlation matrices, and detectability of non-orientable Euclidean manifolds}",
    eprint = "2510.05030",
    archivePrefix = "arXiv",
    primaryClass = "astro-ph.CO",
    reportNumber = "IFT-UAM/CSIC-25-107",
    doi = "10.1088/1475-7516/2026/07/004",
    journal = "JCAP",
    volume = "07",
    pages = "004",
    year = "2026"
}

@article{COMPACT:2024dqe,
    author = "Tamosiunas, Andrius and others",
    collaboration = "COMPACT",
    title = "{Cosmic topology. Part~IVa. Classification of manifolds using machine learning: a case study with small toroidal universes}",
    eprint = "2404.01236",
    archivePrefix = "arXiv",
    primaryClass = "astro-ph.CO",
    reportNumber = "IFT-UAM/CSIC-24-47",
    doi = "10.1088/1475-7516/2024/09/057",
    journal = "JCAP",
    volume = "09",
    pages = "057",
    year = "2024"
}

@software{S3Github,
    collaboration = "COMPACT",
    author = {Martin Barandiaran, Mikel and Petretti, Catherine},
    title = {{CMB-for-S3}},
    url = {https://github.com/CompactCollaboration/CMB-for-S3},
    version = {1.0.0},
    year = {2026}
}

@ARTICLE{Handley2021,
       author = {{Handley}, Will},
        title = "{Curvature tension: Evidence for a closed universe}",
      journal = {Phys. Rev. D.},
         year = 2021,
        month = feb,
       volume = {103},
       number = {4},
          eid = {L041301},
        pages = {L041301},
          doi = {10.1103/PhysRevD.103.L041301},
archivePrefix = {arXiv},
       eprint = {1908.09139},
 primaryClass = {astro-ph.CO},
       adsurl = {https://ui.adsabs.harvard.edu/abs/2021PhRvD.103d1301H}
}

@ARTICLE{DiValentino2020,
       author = {{Di Valentino}, Eleonora and {Melchiorri}, Alessandro and {Silk}, Joseph},
        title = "{Planck evidence for a closed Universe and a possible crisis for cosmology}",
      journal = {Nature Astronomy},
         year = 2020,
        month = feb,
       volume = {4},
        pages = {196-203},
          doi = {10.1038/s41550-019-0906-9},
archivePrefix = {arXiv},
       eprint = {1911.02087},
 primaryClass = {astro-ph.CO},
       adsurl = {https://ui.adsabs.harvard.edu/abs/2020NatAs...4..196D}
}

@article{COMPACT:2026pqs,
    author = "Noltmann, Joline and others",
    collaboration = "COMPACT",
    title = "{Cosmic topology. Part IIc. Detectability with non-standard primordial power spectrum}",
    eprint = "2602.15527",
    archivePrefix = "arXiv",
    primaryClass = "astro-ph.CO",
    reportNumber = "IFT-UAM/CSIC-26-12",
    month = "2",
    year = "2026"
}

@article{Hassfeld:2024hnx,
    author = "Hassfeld, Bjoern and Hebecker, Arthur",
    title = "{The boundary proposal}",
    eprint = "2403.18892",
    archivePrefix = "arXiv",
    primaryClass = "hep-th",
    doi = "10.1016/j.physletb.2024.138946",
    journal = "Phys. Lett. B",
    volume = "856",
    pages = "138946",
    year = "2024"
}

@article{COMPACT:2026nez,
    author = "Mihaylov, Deyan P. and others",
    collaboration = "COMPACT",
    title = "{Cosmic topology. Part Va. Information content of the observable Universe}",
    eprint = "2608.25033",
    archivePrefix = "arXiv",
    primaryClass = "astro-ph.CO",
    month = "8",
    year = "2026"
}

@article{Lewis:2026mif,
    author = "Lewis, Antony",
    title = "{CAMB v2: cosmological power spectra for high-precision surveys}",
    eprint = "2607.14854",
    archivePrefix = "arXiv",
    primaryClass = "astro-ph.CO",
    month = "7",
    year = "2026"
}

@article{Abbott:1986ct,
    author = "Abbott, L. F. and Schaefer, Robert K.",
    title = "{A General, Gauge Invariant Analysis of the Cosmic Microwave Anisotropy}",
    reportNumber = "LHEA-86-003",
    doi = "10.1086/164525",
    journal = "Astrophys. J.",
    volume = "308",
    pages = "546",
    year = "1986"
}

@article{Hu:1997mn,
    author = "Hu, Wayne and Seljak, Uros and White, Martin J. and Zaldarriaga, Matias",
    title = "{A complete treatment of CMB anisotropies in a FRW universe}",
    eprint = "astro-ph/9709066",
    archivePrefix = "arXiv",
    reportNumber = "IASSNS-AST-97-48",
    doi = "10.1103/PhysRevD.57.3290",
    journal = "Phys. Rev. D",
    volume = "57",
    pages = "3290--3301",
    year = "1998"
}

@article{COMPACT:2024cud,
    author = "Samandar, Amirhossein and others",
    collaboration = "COMPACT",
    title = "{Cosmic topology. Part IIIa. Microwave background parity violation without parity-violating microphysics}",
    eprint = "2407.09400",
    archivePrefix = "arXiv",
    primaryClass = "astro-ph.CO",
    reportNumber = "IFT-UAM/CSIC-24-104",
    doi = "10.1088/1475-7516/2024/11/020",
    journal = "JCAP",
    volume = "11",
    pages = "020",
    year = "2024"
}

@article{Lewis:1999bs,
    author = "Lewis, Antony and Challinor, Anthony and Lasenby, Anthony",
    title = "{Efficient computation of CMB anisotropies in closed FRW models}",
    eprint = "astro-ph/9911177",
    archivePrefix = "arXiv",
    doi = "10.1086/309179",
    journal = "Astrophys. J.",
    volume = "538",
    pages = "473--476",
    year = "2000"
}

@article{Planck:2018vyg,
    author = "Aghanim, N. and others",
    collaboration = "Planck",
    title = "{Planck 2018 results. VI. Cosmological parameters}",
    eprint = "1807.06209",
    archivePrefix = "arXiv",
    primaryClass = "astro-ph.CO",
    doi = "10.1051/0004-6361/201833910",
    journal = "Astron. Astrophys.",
    volume = "641",
    pages = "A6",
    year = "2020",
    note = "[Erratum: Astron. Astrophys. 652, C4 (2021)]"
}

@article{Aurich:2012jc,
    author = "Aurich, Ralf and Lustig, Sven",
    title = "{Cosmic Topology of Polyhedral Double-Action Manifolds}",
    eprint = "1207.4378",
    archivePrefix = "arXiv",
    primaryClass = "astro-ph.CO",
    doi = "10.1088/0264-9381/29/23/235028",
    journal = "Class. Quant. Grav.",
    volume = "29",
    pages = "235028",
    year = "2012"
}

@article{Blas:2011rf,
      author         = "Blas, Diego and Lesgourgues, Julien and Tram, Thomas",
      title          = "{The Cosmic Linear Anisotropy Solving System (CLASS) II:
                        Approximation schemes}",
      journal        = "JCAP",
      volume         = "1107",
      year           = "2011",
      pages          = "034",
      doi            = "10.1088/1475-7516/2011/07/034",
      eprint         = "1104.2933",
			note 					= {arXiv:1104.2933},
      archivePrefix  = "arXiv",
      primaryClass   = "astro-ph.CO",
      reportNumber   = "CERN-PH-TH-2011-082, LAPTH-010-11",
      SLACcitation   = "%%CITATION = ARXIV:1104.2933;%%"
}

@article{COMPACT:2025tqc,
    author = "Samandar, Amirhossein and others",
    collaboration = "COMPACT",
    title = "{Cosmic topology. Part IIIb. Eigenmodes and correlation matrices of spin-2 perturbations in orientable Euclidean manifolds}",
    eprint = "2503.08671",
    archivePrefix = "arXiv",
    primaryClass = "astro-ph.CO",
    reportNumber = "IFT-UAM/CSIC-25-25",
    doi = "10.1088/1475-7516/2025/08/015",
    journal = "JCAP",
    volume = "08",
    pages = "015",
    year = "2025"
}

@article{COMPACT:2026uzi,
    author = "Copi, Craig J. and others",
    collaboration = "COMPACT",
    title = "{The topology of the Universe}",
    eprint = "2606.24886",
    archivePrefix = "arXiv",
    primaryClass = "astro-ph.CO",
    doi = "10.1038/s41550-026-02930-6",
    journal = "Nature Astron.",
    volume = "10",
    number = "10",
    pages = "1437--1447",
    year = "2026"
}

@article{COMPACT:2026nfm,
    author = "Samandar, Amirhossein and others",
    collaboration = "COMPACT",
    title = "{Beyond the Kullback--Leibler divergence: the likelihood-ratio distribution as a detectability measure}",
    eprint = "2610.17186",
    archivePrefix = "arXiv",
    primaryClass = "astro-ph.CO",
    reportNumber = "IFT-UAM/CSIC-26-150",
    month = "10",
    year = "2026"
}

@article{Handley:2019anl,
    author = "Handley, Will",
    title = "{Primordial power spectra for curved inflating universes}",
    eprint = "1907.08524",
    archivePrefix = "arXiv",
    primaryClass = "astro-ph.CO",
    doi = "10.1103/PhysRevD.100.123517",
    journal = "Phys. Rev. D",
    volume = "100",
    number = "12",
    pages = "123517",
    year = "2019"
}

@article{Costa:1999ht,
    author = "Costa, S. S. and Fagundes, H. V.",
    title = "{On the birth of a closed hyperbolic universe}",
    eprint = "gr-qc/9911110",
    archivePrefix = "arXiv",
    doi = "10.1023/A:1012236709611",
    journal = "Gen. Rel. Grav.",
    volume = "33",
    pages = "1489--1494",
    year = "2001"
}

@article{Fagundes:2006zy,
    author = "Fagundes, Helio V. and Vargas, Teofilo",
    title = "{Orbifolds, Quantum Cosmology, and Nontrivial Topology}",
    eprint = "gr-qc/0611048",
    archivePrefix = "arXiv",
    month = "11",
    year = "2006"
}

@article{Linde:2004nz,
    author = "Linde, Andrei D.",
    title = "{Creation of a compact topologically nontrivial inflationary universe}",
    eprint = "hep-th/0408164",
    archivePrefix = "arXiv",
    doi = "10.1088/1475-7516/2004/10/004",
    journal = "JCAP",
    volume = "10",
    pages = "004",
    year = "2004"
}

@article{Starobinsky:1993yx,
    author = "Starobinsky, Alexei A.",
    title = "{New restrictions on spatial topology of the universe from microwave background temperature fluctuations}",
    eprint = "gr-qc/9305019",
    archivePrefix = "arXiv",
    reportNumber = "LANDAU-93-04-GR-1",
    journal = "JETP Lett.",
    volume = "57",
    pages = "622--625",
    year = "1993"
}

@article{PhysRevD.42.2458,
  title = {Real tunneling geometries and the large-scale topology of the universe},
  author = {Gibbons, G. W. and Hartle, J. B.},
  journal = {Phys. Rev. D},
  volume = {42},
  issue = {8},
  pages = {2458--2468},
  numpages = {0},
  year = {1990},
  month = {Oct},
  publisher = {American Physical Society},
  doi = {10.1103/PhysRevD.42.2458},
  url = {https://link.aps.org/doi/10.1103/PhysRevD.42.2458}
}

@article{COMPACT:2023rkp,
    author = "Eskilt, Johannes R. and others",
    collaboration = "COMPACT",
    title = "{Cosmic topology. Part~IIa. Eigenmodes, correlation matrices, and detectability of orientable~Euclidean manifolds}",
    eprint = "2306.17112",
    archivePrefix = "arXiv",
    primaryClass = "astro-ph.CO",
    reportNumber = "IFT-UAM/CSIC-23-75",
    doi = "10.1088/1475-7516/2024/03/036",
    journal = "JCAP",
    volume = "03",
    pages = "036",
    year = "2024"
}

@article{COMPACT:2024qni,
    author = "Saha, Samanta and others",
    collaboration = "COMPACT",
    title = "{Cosmic topology. Part Ic. Limits on lens spaces from circle searches}",
    eprint = "2409.02226",
    archivePrefix = "arXiv",
    primaryClass = "astro-ph.CO",
    reportNumber = "IFT-UAM/CSIC-24-118",
    doi = "10.1088/1475-7516/2025/01/004",
    journal = "JCAP",
    volume = "01",
    pages = "004",
    year = "2025"
}

@article{COMPACT:2026vdj,
    author = "Negro, Anna and others",
    collaboration = "COMPACT",
    title = "{Quantum Signatures of Cosmic Topology: How Casimir Backreaction Transmits Isotropy Violation}",
    eprint = "2603.12319",
    archivePrefix = "arXiv",
    primaryClass = "hep-th",
    reportNumber = "IFT-UAM/CSIC-26-25",
    month = "3",
    year = "2026"
}

@article{Negro:2026ooh,
    author = "Negro, Anna and Hinterbichler, Kurt and Starkman, Glenn D. and Akrami, Yashar and Anselmi, Stefano and Carr{\'o}n Duque, Javier and Martin Barandiaran, Mikel and Tamosiunas, Andrius",
    collaboration = "COMPACT",
    title = "{Breaking de Sitter Symmetry with Cosmic Topology: Cosmological Role of Infrared Vacuum Freedom}",
    eprint = "2609.28637",
    archivePrefix = "arXiv",
    primaryClass = "hep-th",
    month = "9",
    year = "2026"
}

@article{kullback1951,
author = {S. Kullback and R. A. Leibler},
title = {{On Information and Sufficiency}},
volume = {22},
journal = "Ann. Math. Stat.",
number = {1},
publisher = {Institute of Mathematical Statistics},
pages = {79 -- 86},
year = {1951},
doi = {10.1214/aoms/1177729694},
URL = {https://doi.org/10.1214/aoms/1177729694}
}

@book{kullback1959information,
  title={Information Theory and Statistics},
  author={Kullback, S.},
  lccn={59005881},
  series={Wiley publication in mathematical statistics},
  url={https://books.google.com/books?id=XeRQAAAAMAAJ},
  year={1959},
  publisher={Wiley}
}

@article{Kiefer:2021iko,
    author = "Kiefer, Claus and Vardanyan, Tatevik",
    title = "{Power spectrum for perturbations in an inflationary model for a closed universe}",
    eprint = "2111.07835",
    archivePrefix = "arXiv",
    primaryClass = "gr-qc",
    doi = "10.1007/s10714-022-02918-3",
    journal = "Gen. Rel. Grav.",
    volume = "54",
    number = "4",
    pages = "30",
    year = "2022"
}

@article{Vardanyan:2023ldj,
    author = "Vardanyan, Tatevik and Kiefer, Claus",
    title = "{Quantum-gravitational corrections to the power spectrum for a closed universe}",
    eprint = "2302.07001",
    archivePrefix = "arXiv",
    primaryClass = "gr-qc",
    doi = "10.1088/1742-6596/2533/1/012036",
    journal = "J. Phys. Conf. Ser.",
    volume = "2533",
    number = "1",
    pages = "012036",
    year = "2023"
}

@article{Lesgourgues:2011re,
    author = "Lesgourgues, Julien",
    title = "{The Cosmic Linear Anisotropy Solving System (CLASS) I: Overview}",
    eprint = "1104.2932",
    archivePrefix = "arXiv",
    primaryClass = "astro-ph.IM",
    month = "4",
    year = "2011"
}
\end{document}